\documentclass[twocolumn]{aastex631}
\usepackage[utf8]{inputenc}
\usepackage{amsmath}
\usepackage{amssymb}
\usepackage{graphicx}
\usepackage{xcolor}
\usepackage{enumitem}

\received{11 Jun 2024}
\revised{---}
\accepted{---}

\submitjournal{ApJ}

\shorttitle{The Flip Side of Mass Loading}
\shortauthors{Voit et al.}

\begin{document}

\title{\textbf{Equilibrium States of Galactic Atmospheres I: The Flip Side of Mass Loading}}


\author[0000-0002-3514-0383]{G. Mark Voit}
\affiliation{Michigan State University,
Department of Physics and Astronomy,
East Lansing, MI 48824, USA}

\author[0000-0002-2499-9205]{Viraj Pandya}
\altaffiliation{Hubble Fellow}
\affiliation{Columbia University,
Department of Astronomy,
New York, NY 10027, USA}

\author[0000-0003-3806-8548]{Drummond B. Fielding}
\affiliation{Center for Computational Astrophysics, Flatiron Institute, New York, NY 10010, USA}
\affiliation{Department of Astronomy, Cornell University, Ithaca, NY 14853, USA}

\author[0000-0003-2630-9228]{Greg L. Bryan}
\affiliation{Columbia University,
Department of Astronomy,
New York, NY 10027, USA}

\author[0000-0002-5840-0424]{Christopher Carr}
\affiliation{Columbia University,
Department of Astronomy,
New York, NY 10027, USA}

\author[0000-0002-2808-0853]{Megan Donahue}
\affiliation{Michigan State University,
Department of Physics and Astronomy,
East Lansing, MI 48824, USA}

\author[0000-0003-4754-6863]{Benjamin D. Oppenheimer}
\affiliation{University of Colorado, 
Center for Astrophysics and Space Astronomy, 389 UCB, Boulder, CO 80309, USA}

\author[0000-0002-6748-6821]{Rachel S. Somerville}
\affiliation{Center for Computational Astrophysics,
Flatiron Institute,
New York, NY 10010, USA}



\begin{abstract}
    This paper presents a new framework for understanding the relationship between a galaxy and its circumgalactic medium (CGM). It focuses on how \textit{imbalances} between heating and cooling cause either expansion or contraction of the CGM. It does this by tracking \textit{all} of the mass and energy associated with a halo's baryons, including their gravitational potential energy, even if feedback has pushed some of those baryons beyond the halo's virial radius. We show how a star-forming galaxy's equilibrium state can be algebraically derived within the context of this framework, and we analyze how the equilibrium star formation rate depends on supernova feedback. We consider the consequences of varying the mass loading parameter $\eta_M \equiv \dot{M}_{\rm wind} / \dot{M}_*$ relating a galaxy's gas mass outflow rate ($\dot{M}_{\rm wind}$) to its star formation rate ($\dot{M}_*$) and obtain results that challenge common assumptions. In particular, we find that equilibrium star formation rates in low-mass galaxies are generally insensitive to mass loading, and when mass loading does matter, increasing it actually results in \textit{more} star formation because more supernova energy is needed to resist atmospheric contraction.
\end{abstract}

\section{Introduction}
\label{sec:Introduction}

You are about to read a paper that radically oversimplifies the phenomenon of galaxy evolution. Where you might expect to see a cutting edge numerical simulation or a sophisticated semi-analytic model, you will instead find a system of three ordinary differential equations, sometimes reduced to just two. This radical reduction in complexity makes it possible to derive algebraic solutions defining a star-forming galaxy's equilibrium state and clarifying how a galaxy's time-averaged star formation rate depends on the properties of supernova feedback. Our goal is therefore complementary to the goals of cosmological numerical simulations and complex semi-analytic models: We are trying to provide simple conceptual tools for interpreting and comparing simulations, models, and observational data sets that are far more complicated.

This paper is the third in a series starting with \citet{Carr_2023ApJ...949...21C} and continuing through \citet{Pandya_2023ApJ...956..118P}. All three present regulator models for galaxy evolution that emphasize the role of the circumgalactic medium (CGM) and its energy content in governing the supply of star-forming gas. You can find more extensive discussion of the motivations for the series in the first two papers. A companion paper (Voit et al. 2024, hereafter Paper II) relates this paper's models to previous semi-analytic models for the CGM and suggests how those models may be used to understand differences between cosmological numerical simulations of galaxy evolution.

The new regulator model differs from the previous two, and from many other semi-analytic models for galaxy evolution, in two key respects:
\begin{enumerate}
    \item It applies a comprehensive accounting approach to the \textit{total energy} of a halo's baryons, including the energy content of baryons that feedback may have pushed far beyond the halo's virial radius.
    \item It explicitly includes gravitational potential energy in the baryonic energy budget, because imbalances between energy input and radiative losses ultimately cause either expansion or contraction of the atmosphere surrounding a galaxy.
\end{enumerate}
Another difference from many other semi-analytic models, including \citet{Pandya_2023ApJ...956..118P} but not \citet{Carr_2023ApJ...949...21C}, is that the new model allows the CGM to supply a galaxy with gas even if feedback energy input exceeds radiative cooling of the galaxy's atmosphere. 

We will show that our approach helps to clarify some misconceptions about the relationship between galactic winds and the fraction $f_*$ of a halo's baryons that become stars. Early discussions of that relationship \citep[e.g.,][]{Larson_1974MNRAS.169..229L,ChevalierClegg1985Natur.317...44C,DekelSilk1986ApJ...303...39D, WhiteFrenk1991ApJ...379...52W} suggested to many that galactic winds might limit star formation by permanently ejecting a galaxy's gas. Much of the published work on galactic winds has therefore focused on the \textit{mass loading parameter} $\eta_M \equiv \dot{M}_{\rm wind} / \dot{M}_*$ relating a galaxy's gas mass outflow rate ($\dot{M}_{\rm wind}$) to its star formation rate ($\dot{M}_*$). If most of the baryons accreting onto a halo end up in its central galaxy, then $\eta_M$ needs to depend strongly on halo mass ($M_{\rm halo}$) in order to explain observations of the $f_*$--$M_{\rm halo}$ relation indicating that $f_*$ is nearly an order of magnitude smaller at $M_{\rm halo} \sim 10^{11} \, M_\odot$ than at $M_{\rm halo} \sim 10^{12} \, M_\odot$ \citep[e.g.,][]{McGaugh+2010ApJ...708L..14M,Moster_2010ApJ...710..903M,Behroozi2019MNRAS.488.3143B}. 

Some simulations produce dwarf galaxies with outflows having $\eta_M \gtrsim 10$ \citep[e.g.,][]{Vogelsberger_IllustrisModel_2013MNRAS.436.3031V,Muratov_2015MNRAS.454.2691M,Pandya_2021MNRAS.508.2979P}, but observational constraints on $\eta_M$ usually indicate smaller proportions of mass loading \citep[e.g.,][]{Martin_1999ApJ...513..156M,Heckman_2015ApJ...809..147H,Chisholm_2017MNRAS.469.4831C,McQuinn_2019ApJ...886...74M}. Similarly, simulations that focus on the galactic disk and resolve individual supernova events tend to find lower mass loading values \citep{SILCC2016, Fielding2018, LiBryan2020, Kim2020} Also, semi-analytic models that rely solely on gas ejection, as parameterized by $\eta_M$, tend to have trouble reproducing the mass-metallicity relationships observed among low-mass galaxies \citep[e.g.,][]{SomervilleDave_2015ARA&A..53...51S}. Those tensions have prompted questions about how much of a halo's gas actually cycles through its central galaxy and have stimulated interest in preventative feedback modes that reduce a galaxy's gas supply \citep[e.g.,][]{Oppenheimer_2010MNRAS.406.2325O,vandeVoort_2011MNRAS.414.2458V,DaveFinlatorOppenheimer_2012MNRAS.421...98D,LuMoWechsler_2015MNRAS.446.1907L,Hirschmann_2016MNRAS.461.1760H,LuBenson_2017ApJ...846...66L,Pandya_2020ApJ...905....4P,MitchellSchaye_2022MNRAS.511.2948M,Carr_2023ApJ...949...21C,Pandya_2023ApJ...956..118P}.

One interesting feature of the feedback models in \citet{Carr_2023ApJ...949...21C} is the insensitivity of their predicted $f_*$--$M_{\rm halo}$ relations to $\eta_M$. This paper explains why \cite{Carr_2023ApJ...949...21C} arrived at that result and outlines the general conditions that can make $f_*$ insensitive to $\eta_M$. It also discusses why large values of $\eta_M$ can have counterintuitive effects, sometimes causing $f_*$ to increase. By presenting and interpreting those findings, both here and in Paper II, we hope to reframe how astronomers view the relationship between supernova feedback and mass loading of the outflows it produces.

Here is how things proceed: The next section (\S \ref{sec:Foundation}) lays the paper's conceptual foundation and previews the findings that follow. Section \ref{sec:Dynamics} then makes the conceptual reasoning more quantitative by introducing a ``minimalist" regulator model designed to represent as simply as possible how supernova-driven galactic outflows shape the global properties of a circumgalactic atmosphere. Section \ref{sec:Asymptotic} analyzes the equilibrium states of that model, which illustrate how convergence of a galactic atmosphere toward a quasi-steady state and the resulting value of $f_*$ depend on the properties of supernova feedback. 
Section \ref{sec:Ballistic} considers the contrasting case of supernova-driven outflows that \textit{do not} interact with the rest of a galaxy's atmosphere. 
The concluding section summarizes our findings. A glossary in the Appendix describes the many symbols introduced along the way.


\section{Conceptual Foundation}
\label{sec:Foundation}

\begin{figure*}[t]
\begin{center}
  \includegraphics[width=0.97\textwidth]{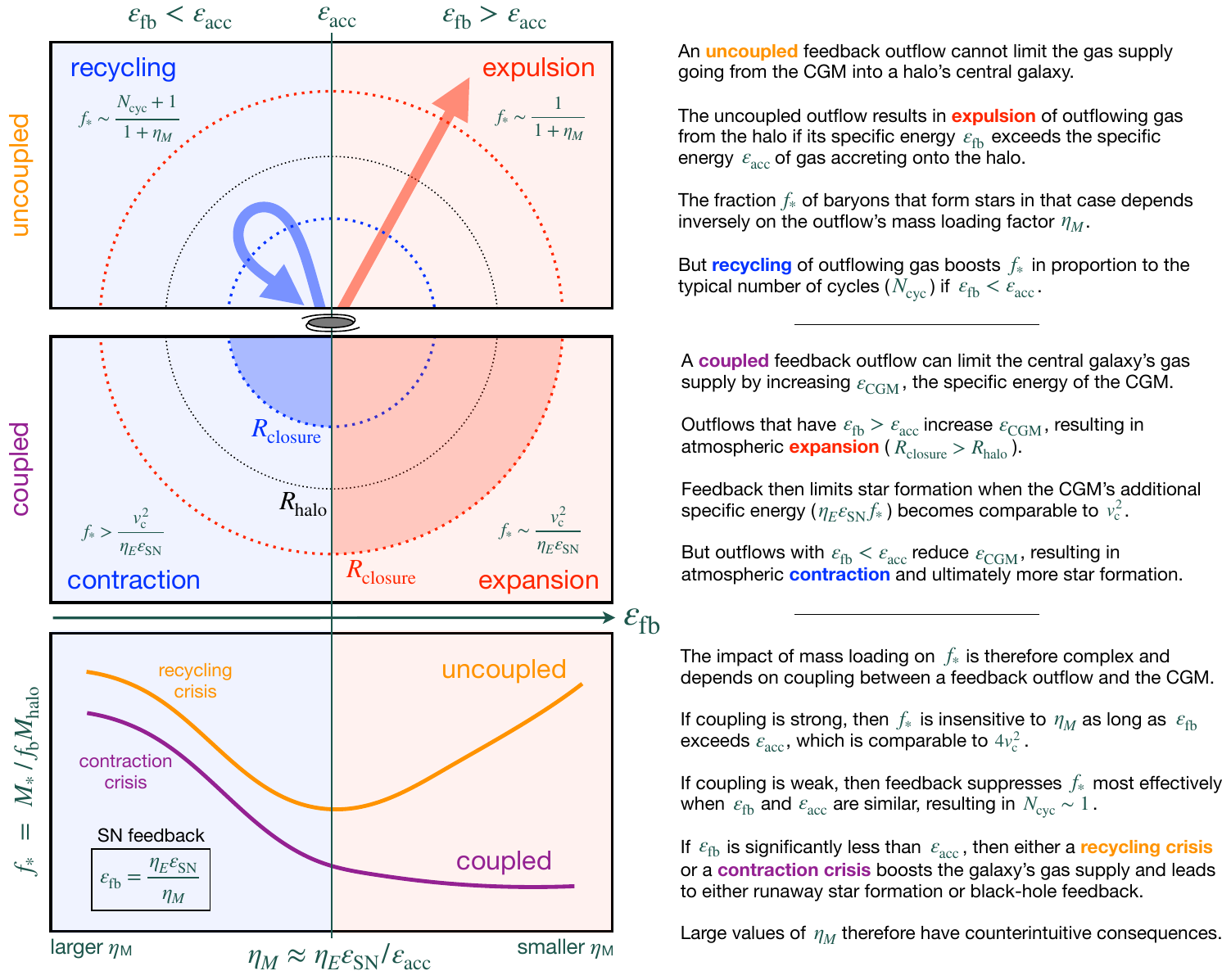}
\end{center}
\caption{Schematic summary of the paper's foundational concepts and key results.
    }
\label{fig:FlipSideGraphic}
\end{figure*}

To prepare readers for the paper's main arguments, we would like to make sure that the following foundational concepts are crystal clear:

\begin{itemize}

    \item \textbf{Specific Energy.} 
    The total specific energy $\varepsilon_{\rm CGM}$ of a galaxy's circumgalactic atmosphere consists of gravitational potential energy, thermal energy, and non-thermal energy.\footnote{The kinetic energy of fluid flow, including both bulk flows and turbulence, is part of the non-thermal category.}

    Specifying the gravitational potential's zero point so that gas entering or leaving a halo's central galaxy has zero potential energy (instead of putting the zero point at infinity) helps to clarify the gravitational potential's role. 
    
    With that zero point, gas accreting at time $t$ onto a cosmological halo with circular velocity $v_{\rm c}$ has a specific energy $\varepsilon_{\rm acc}(t) \approx 4 v_{\rm c}^2$, consisting mostly of gravitational potential energy.\footnote{The factor of four corresponds to a nearly isothermal halo potential well containing a central galaxy having a radius $\sim 50$ times smaller than the halo's virial radius (because $\ln 50 \approx 4$). A comparable factor applies to an NFW potential \citep{nfw97}, which has a depth approximately $4.6$ times its maximum value of $v_{\rm c}^2$.} Halo gas must therefore radiate an amount of specific energy similar to $\varepsilon_{\rm acc}$ before it can descend into the halo's central galaxy and form stars there.

    Feedback produces outflows of gas that have a specific energy $\varepsilon_{\rm fb}$ when they leave the galaxy. Initially, the outflowing gas has zero potential energy, but its thermal and kinetic energy convert into gravitational potential energy as the outflow rises to greater altitudes. 
    
    Figure \ref{fig:FlipSideGraphic} schematically summarizes how feedback outcomes depend on the relationship between $\varepsilon_{\rm fb}$ and $\varepsilon_{\rm acc}$. Cases with $\varepsilon_{\rm fb} < \varepsilon_{\rm acc}$ are on the left, and cases with $\varepsilon_{\rm fb} > \varepsilon_{\rm acc}$ are on the right.

    \item \textbf{Feedback Coupling.} 
    The fate of a feedback outflow depends on how closely it couples with a galaxy's circumgalactic atmosphere. Here we consider two limiting cases: 
    
    \begin{enumerate}
        \item \textbf{Uncoupled Outflows.} At one extreme are feedback outflows that are \textbf{uncoupled} because they never transfer any of their energy to the rest of the CGM and therefore cannot reduce the supply of gas going from the CGM into a halo's central galaxy. The top panel of Figure \ref{fig:FlipSideGraphic} represents what happens to uncoupled outflows.

        If the specific energy $\varepsilon_{\rm fb}$ of an uncoupled outflow significantly exceeds $\varepsilon_{\rm acc}(t)$, then the recycling timescale $t_{\rm cyc}$ on which outflowing gas returns to the halo's central galaxy is longer than the cosmic time $t$. That is because $\varepsilon_{\rm acc}(t)$ is the specific energy of a cosmological trajectory that reached its maximum altitude at a cosmic time similar to $t/2$ and completes its descent at time $t$. An outflow with greater specific energy will take even longer to rise to its maximum altitude and then return. The fate of an uncoupled outflow with $\varepsilon_{\rm fb} > \varepsilon_{\rm acc}$ is therefore \textbf{expulsion} from the halo.
    
        However, gas exiting a galaxy in an uncoupled outflow with $\varepsilon_{\rm fb} < \varepsilon_{\rm acc}$ can fall back into the galaxy on a timescale shorter than $t$. In that case, the result of feedback is \textbf{recycling} capable of ejecting gas from a galaxy multiple times.
    
        The number of times $N_{\rm cyc} \sim t / t_{\rm cyc}$ that galactic gas is typically recycled determines how effectively uncoupled outflows suppress galactic star formation. If $N_{\rm cyc} \ll 1$, then the amount of gas that forms stars is $1 / (1 + \eta_M)$ times the gas mass that has passed through the galaxy. But if $N_{\rm cyc}$ is larger, then the proportion of cycling gas that forms stars is $(N_{\rm cyc} + 1) / (1 + \eta_M)$, as long as $N_{\rm cyc} \ll \eta_M$.
    
        Section \ref{sec:Ballistic} presents a more quantitative analysis of uncoupled outflows that supports this line of reasoning.
    
        \item \textbf{Coupled Outflows.} 
        An outflow that transfers energy to the rest of the CGM is considered \textbf{coupled} because  the energy it transfers can reduce the central galaxy's gas supply. 
        
        The rate of energy transfer from a coupled supernova-driven outflow to the CGM can be expressed as $\eta_E \varepsilon_{\rm SN} \dot{M}_*$, in which $\varepsilon_{\rm SN}$ is the specific supernova energy released by a stellar population and $\eta_E$ is an \textit{energy-loading parameter} specifying the fraction of supernova energy transported into the CGM. The specific energy of such a coupled outflow is $\varepsilon_{\rm fb} = \eta_E \varepsilon_{\rm SN}/ \eta_M$.
        
        Sections \ref{sec:Dynamics} and \ref{sec:Asymptotic} present quantitative models for coupled supernova outflows showing that coupled outflows with $\varepsilon_{\rm fb} > \varepsilon_{\rm acc}$ cause \textbf{expansion} of the CGM, while coupled outflows with $\varepsilon_{\rm fb} < \varepsilon_{\rm acc}$ allow \textbf{contraction} of the CGM. 
        
        The middle panel of Figure \ref{fig:FlipSideGraphic} schematically illustrates the results of \S \ref{sec:Dynamics} and \S \ref{sec:Asymptotic} by depicting the relationship between a halo's virial radius $R_{\rm halo}$ and the \textit{closure radius} $R_{\rm closure}$ within which the proportion of mass in baryons equals the cosmic baryon mass fraction $f_{\rm b}$ \citep[see][]{Ayromlou_2023MNRAS.524.5391A}. Expanded atmospheres have $R_{\rm closure} > R_{\rm halo}$, while contracted atmospheres can have $R_{\rm closure} < R_{\rm halo}$.
    
    \end{enumerate}

    \item \textbf{Asymptotic Star Formation.} 
    Each of the four cases illustrated in the upper two panels of Figure \ref{fig:FlipSideGraphic} converges toward an asymptotic star formation rate and stellar baryon fraction that depend on either $\eta_M$ or $\eta_E$, or both. Section \ref{sec:Asymptotic} analyzes the star-forming properties of galaxies with coupled outflows. Section \ref{sec:Ballistic} analyzes the star-forming properties of galaxies with uncoupled outflows.  

    The figure's bottom panel schematically summarizes three primary results of those analyses:
    \begin{enumerate} 

        \item Uncoupled outflows with high specific energy ($\varepsilon_{\rm fb} \gg \varepsilon_{\rm acc}$) lead to stellar baryon fractions that are inversely proportional to $1 + \eta_M$ and insensitive to $\eta_E$. 

        \item Coupled outflows with high specific energy ($\varepsilon_{\rm fb} \gg \varepsilon_{\rm acc}$) lead to stellar baryon fractions that are inversely proportional to $\eta_E$ and insensitive to $\eta_M$. 
        
        \item Both coupled and uncoupled outflows fail to limit star formation if $\eta_M$ is large enough to make $\varepsilon_{\rm fb}$ significantly smaller than $\varepsilon_{\rm acc}$. Uncoupled outflows face a \textbf{recycling crisis} when $t_{\rm cyc}$ becomes much smaller than $t$. Coupled outflows face a \textbf{contraction crisis} when gas flowing out of the central galaxy has a specific energy much less than $\varepsilon_{\rm acc} \approx 4 v_{\rm c}^2$. In both cases, the crisis leads to larger amounts of star formation than a galaxy with $\varepsilon_{\rm fb} \approx \varepsilon_{\rm acc}$ would have.
        
    \end{enumerate}

\end{itemize}
While the supporting calculations in sections \ref{sec:Dynamics} through \ref{sec:Ballistic} are highly idealized compared to sophisticated cosmological simulations of galaxy evolution, we believe they provide important insights into the outcomes of those simulations, for reasons explained in Paper II.

\section{Coupled Outflows: Regulator Model}
\label{sec:Dynamics}

We will start by examining how coupled outflows self-regulate. The first step is to develop a regulator model that mimics the interplay between a galaxy's star formation rate and its gas supply \citep[e.g.,][]{Bouche_2010ApJ...718.1001B,DaveFinlatorOppenheimer_2012MNRAS.421...98D,Lilly+2013ApJ...772..119L}. 

\subsection{Comprehensive Accounting}

The regulator model presented here keeps track of all of the baryons that have entered a cosmological halo. Accretion brings baryons into the halo at the rate $\dot{M}_{\rm acc}$. Some of that mass ends up in stars ($M_*$). Some stays in the interstellar medium ($M_{\rm ISM}$). The rest remains circumgalactic. A comprehensive accounting approach to the baryon budget therefore defines the mass of the central galaxy's circumgalactic medium (CGM) to be 
\begin{equation}
    M_{\rm CGM} = M_{\rm acc} - M_* - M_{\rm ISM}
    \label{eq:MassBudget}
    \; \; .
\end{equation}
Unlike many previous approaches, there is no term representing ejection of gas mass from the CGM. Accreted gas belongs either to the galaxy or its CGM and is never lost from the system. However, feedback from coupled outflows can push some of the CGM beyond $R_{\rm halo}$, causing $R_{\rm closure}$ to exceed $R_{\rm halo}$.

Energy associated with the gas mass $M_{\rm CGM}$ can be either gravitational potential energy ($E_\varphi$), thermal energy ($E_{\rm th}$), or non-thermal energy ($E_{\rm nt}$). The total energy content of the circumgalactic gas is therefore
\begin{equation}
    E_{\rm CGM} = E_\varphi + E_{\rm th} + E_{\rm nt} 
    \label{eq:EnergyBudget}
    \; \; .
\end{equation}
Adding either thermal or non-thermal energy generally causes the circumgalactic medium to expand. Expansion then converts the added energy into gravitational potential energy. Likewise, losses of energy generally permit atmospheric contraction that converts gravitational potential energy into either thermal or non-thermal energy.

\subsection{Energy Sources and Sinks}

A circumgalactic atmosphere's total energy evolves according to
\begin{equation}
    \dot{E}_{\rm CGM} = \dot{E}_{\rm acc} - \dot{E}_{\rm rad} - \dot{E}_{\rm in} + \dot{E}_{\rm fb} + \dot{E}_\varphi 
    \label{eq:EnergySourcesSinks}
    \; \; .
\end{equation}
Cosmological accretion adds energy at the rate $\dot{E}_{\rm acc}$. Radiative losses ($\dot{E}_{\rm rad}$) allow some of the accreted gas to sink into the central galaxy. Gas entering the central galaxy's interstellar medium removes energy from the circumgalactic medium at the rate $\dot{E}_{\rm in}$. Star formation fueled by interstellar gas then produces a feedback response that adds energy to the circumgalactic atmosphere at the rate $\dot{E}_{\rm fb}$. While all that is happening, the gravitational potential well confining the circumgalactic gas evolves. The last source term 
\begin{equation}
    \dot{E}_\varphi \equiv \int \dot{\varphi} \, dM_{\rm CGM}
    \label{eq:Edot_phi}
\end{equation}
represents the change in atmospheric energy coming solely from time-dependent changes in the gravitational potential $\varphi$.

\subsection{Gravitational Potential}

To simplify later calculations, we will approximate the halo's gravitational potential as a singular isothermal sphere:
\begin{equation}
    \varphi (r) = v_{\rm c}^2 \ln \left( \frac {r} {r_0} \right)
    \label{eq:SIS_Potential}
    \; \; .
\end{equation}
This potential well has a constant circular velocity $v_{\rm c}$ and a zero point at $r = r_0$. We will assume that $v_{\rm c}$ and $r_0$ both remain constant with time, so that $\dot{E}_\varphi = 0$. 

Gas accreting through the virial radius $R_{\rm halo}$ of a galaxy's cosmological halo at the rate $\dot{M}_{\rm acc}$ has a specific energy $\varepsilon_{\rm acc} \equiv \dot{E}_{\rm acc} / \dot{M}_{\rm acc}$ at least as large as $\varphi(R_{\rm halo})$. Before accreting gas can enter the central galaxy, it needs to lose some of that energy. Here we will choose $r_0$ to represent the inner boundary of the circumgalactic medium, so that gas passing from the circumgalactic medium into the interstellar medium has zero potential energy. This choice makes $\varphi (R_{\rm halo})$ similar to the specific energy that accreting gas needs to lose in order to enter the central galaxy and remain within it. The magnitude of $\varphi (R_{\rm halo})$ ranges from $3.9 v_{\rm c}^2$ to $4.6 v_{\rm c}^2$ as $R_{\rm halo}$ ranges from $50 r_0$ to $100 r_0$ in an isothermal potential well.

\subsection{Mass Exchange}

Suppose that circumgalactic radiative cooling allows atmospheric gas to enter the central galaxy at the rate $\dot{M}_{\rm in}$ and that feedback from star formation expels gas from the central galaxy at a rate that is $\eta_M$ times the star formation rate $\dot{M}_*$. The interstellar gas mass then evolves (neglecting mass return from stars for simplicity) according to 
\begin{equation}
    \dot{M}_{\rm ISM} 
        = \dot{M}_{\rm in} - \left( 1 + \eta_M \right) \dot{M}_*
        \label{eq:Mdot_ISM}
    \; \; ,
\end{equation}
and the circumgalactic gas mass evolves according to 
\begin{equation}
    \dot{M}_{\rm CGM} 
        = \dot{M}_{\rm acc} 
            - \dot{M}_{\rm in} 
            + \eta_M \dot{M}_*
                    \label{eq:Mdot_CGM}
        \; \; .
\end{equation}
We will soon be paying closer attention to the mass loading parameter $\eta_M$ that appears in both of these equations.

\subsection{Supernova Feedback}

Star formation in the central galaxy produces a feedback response that can limit further star formation in two distinct ways. First, it can eject gas from the galaxy, as represented by the mass loading parameter $\eta_M$. Second, it can add energy to the circumgalactic medium, causing it to expand, thereby reducing the galaxy's gas supply $\dot{M}_{\rm in}$. Supernova feedback energy enters the galaxy's atmosphere at a rate
\begin{equation}
    \dot{E}_{\rm fb} = \eta_E \varepsilon_{\rm SN} \dot{M}_*
\end{equation}
in which the specific kinetic energy output from a typical stellar population is
\begin{equation}
    \varepsilon_{\rm SN} \approx \frac {10^{51} \, {\rm erg}} {100 \, M_\odot} \approx \left( 700 \, {\rm km \, s^{-1}} \right)^2
    \; \; .
\end{equation}
The energy-loading parameter $\eta_E$ specifies the proportion of supernova energy reaching the circumgalactic medium and coupling with it.

\subsection{Radiative Losses}

For expansion to be effective, the supernova energy coupling with the galaxy's atmosphere ($\dot{E}_{\rm fb}$) needs to be at least comparable to the atmosphere's radiative loss rate ($\dot{E}_{\rm rad}$). We can relate the atmosphere's radiative losses to the galaxy's gas supply via the specific energy
\begin{equation}
    \varepsilon_{\rm rad} \equiv \frac {\dot{E}_{\rm rad}}
    {\dot{M}_{\rm in}} 
        \label{eq:eps_rad}
    \; \; .
\end{equation}
This quantity will turn out to be useful when we compare $\dot{E}_{\rm fb}$ with $\dot{E}_{\rm rad}$. Under most circumstances, we expect $\varepsilon_{\rm rad}$ to be similar to $\varepsilon_{\rm acc} \approx \varphi (R_{\rm halo})$, because the latter quantity is the specific potential energy that accreted gas needs to lose in order to get into the central galaxy. For completeness, we also define
\begin{equation}
    \varepsilon_{\rm in} 
        \equiv \frac {\dot{E}_{\rm in}} 
                    {\dot{M}_{\rm in}} 
        \label{eq:eps_in}
\end{equation}
and 
\begin{equation}
    \varepsilon_{\rm loss} 
        \equiv \varepsilon_{\rm rad} + \varepsilon_{\rm in}
        \label{eq:eps_loss}
\end{equation} 
so that the \textit{total} rate of atmospheric energy loss becomes $\dot{E}_{\rm loss} = \dot{M}_{\rm in} \varepsilon_{\rm loss}$.

\subsection{Net Feedback Energy}

Suppose we now integrate atmospheric energy losses and gains over a galaxy's entire history, to capture both the radiative losses needed to make a stellar mass $M_*$ and the resulting feedback response. The cumulative amount of supernova feedback energy added to the atmosphere by coupled outflows is $E_{\rm fb} = \eta_E \varepsilon_{\rm SN} M_*$. The cumulative radiative losses follow from integrating $\dot{E}_{\rm rad} = \dot{M}_{\rm in} \varepsilon_{\rm rad}$ over time, giving
\begin{equation}
    E_{\rm rad} 
        = \left[ \langle 1 + \eta_M \rangle M_* + M_{\rm ISM} \right]
          \langle \varepsilon_{\rm rad} \rangle
          \label{eq:Erad}
          \; \; ,
\end{equation}
in which
\begin{equation}
    \left\langle 1 + \eta_M \right\rangle 
        \equiv \frac {1} {M_*}
                \int (1 + \eta_M) \dot{M}_* \, dt 
                     \; \; .
\end{equation}
and 
\begin{equation}
    \left\langle \varepsilon_{\rm rad} \right\rangle 
        \equiv \frac {\int \dot{M}_{\rm in} \varepsilon_{\rm rad} \, dt} 
                     {\int \dot{M}_{\rm in} \, dt}
    \; \; .
\end{equation}
The net effect of cooling and feedback on the total energy content of a galaxy's atmosphere is therefore
\begin{align}
    E_{\rm fb} - E_{\rm rad} 
        &= \left[ \eta_E \varepsilon_{\rm SN} 
            - \langle 1 + \eta_M \rangle
              \left\langle \varepsilon_{\rm rad} \right\rangle
              \right]
              M_* 
            \nonumber \\
        &\quad\quad - \left\langle \varepsilon_{\rm rad} \right\rangle
              M_{\rm ISM}
    \label{eq:NetFeedbackEnergy}
    \; \; .
\end{align}
Meanwhile, the gas supply fueling galactic star formation transfers a total energy
\begin{equation}
    E_{\rm in} 
        = \left[ \langle 1 + \eta_M \rangle M_* + M_{\rm ISM}
              \right]
              \left\langle \varepsilon_{\rm in} \right\rangle
\end{equation}
from the galaxy's circumgalactic medium to its interstellar medium. Here, $\langle \varepsilon_{\rm in} \rangle$ is a mass-averaged value of $\varepsilon_{\rm in}$ similar to $\langle \varepsilon_{\rm rad} \rangle$.

\subsection{The Flip Side of Mass Loading}
\label{sec:FlipSide}

Now we revisit the mass loading parameter $\eta_M$. 
Its most obvious effect on star formation in the central galaxy is to suppress it by ejecting potentially star forming gas from the galaxy's ISM. However, the presence of $\eta_M$ following a minus sign in \autoref{eq:NetFeedbackEnergy}, the formula for net feedback energy, suggests another side to the mass loading story.

According to our energy accounting exercise, increasing $\eta_M$ \textit{reduces} the supernova feedback cycle's net energy input into a galaxy's atmosphere. In fact, a supernova feedback cycle with 
\begin{equation}
    \langle 1 + \eta_M \rangle 
    \langle \varepsilon_{\rm rad} \rangle 
        > \eta_E \varepsilon_{\rm SN}
\end{equation}
results in a net atmospheric energy \textit{loss} rather than a net energy gain.\footnote{This condition is sufficient but not necessary, because it omits, for simplicity, the additional energy loss terms proportional to $M_{\rm ISM}$ and $\langle \varepsilon_{\rm in} \rangle$.} 

\textbf{\textit{This underappreciated feature of mass-loaded outflows---that more mass loading implies greater atmospheric radiative losses---is the key point of the entire paper.}} It is consequential because a galaxy's atmosphere responds to energy loss by contracting and becoming denser, possibly increasing the galaxy's gas supply, its star formation rate, and its total stellar mass. The combined effects of $\eta_M$ on $M_*$ are therefore interestingly coupled and require closer examination. 

\subsection{Minimalist Regulator Model}
\label{sec:Minimalist}

Section \ref{sec:Asymptotic} will explore the impact of mass loaded galactic outflows on galaxies and their atmospheres in the context of a highly simplified evolutionary model. Three coupled differential equations define the model:
\begin{align}
    \dot{E}_{\rm CGM} \:
        &= \: \dot{M}_{\rm acc} \varepsilon_{\rm acc}
                \: - \: \dot{M}_{\rm in} 
                    \varepsilon_{\rm loss}
                \: + \: \eta_E \varepsilon_{\rm SN}   
                    \frac {M_{\rm ISM}} {t_{\rm SF}}
            \\
    \dot{M}_{\rm CGM} \:
        &= \: \dot{M}_{\rm acc}
                \: - \: \dot{M}_{\rm in} 
                \: + \: \eta_M 
                    \frac {M_{\rm ISM}} {t_{\rm SF}}            
            \\
    \dot{M}_{\rm ISM} \:
        &= \: \dot{M}_{\rm in}
                \: - \: \left( 1 + \eta_M \right) 
            \frac {M_{\rm ISM}} {t_{\rm SF}}  
            \; \; .
\end{align}
We will call this system of equations the \textit{minimalist regulator model}. For a given halo, both $\dot{M}_{\rm acc}$ and $\varepsilon_{\rm acc}$ are cosmological source terms that depend only on time. The galaxy's star formation timescale $t_{\rm SF} \equiv M_{\rm ISM} / \dot{M}_*$ can be either constant, or a function of time, or a function of both time and $M_{\rm ISM}$. Both the galaxy's gas supply $\dot{M}_{\rm in}$ and the energy loss parameter $\varepsilon_{\rm loss}$ are presumed to be functions of $E_{\rm CGM}$ and $M_{\rm CGM}$. They may also depend explicitly on time.

Notice that the main effect of the $M_{\rm ISM}$ equation is to spread the star formation fueled by incoming gas over a timescale $\sim t_{\rm SF} / (1 + \eta_M)$. Integrating the $M_{\rm ISM}$ equation using the integrating factor method gives
\begin{equation}
    M_{\rm ISM} (t)
        \: = \: \int_0^t
            \dot{M}_{\rm in} (t^\prime)
            \, \exp \left[ \tau(t^\prime) 
                        - \tau(t) \right]
            \, dt^\prime
\end{equation}
in which  
\begin{equation}
    \tau (t) 
     \equiv \int_0^t \frac {1+\eta_M}
                        {t_{\rm SF} (t^\prime)}
                        \, dt^\prime
    \\[5pt]
\end{equation}
is a gas depletion parameter defined so that the exponential factor represents the fraction of gas that entered the galaxy at time $t^\prime$ and still remains in the interstellar medium at time $t$. The galaxy's star formation rate is therefore $\dot{M}_* = \langle \dot{M}_{\rm in} \rangle / (1 + \eta_M)$, in which $\langle \dot{M}_{\rm in} \rangle \equiv (1 + \eta_M) M_{\rm ISM} / t_{\rm SF}$ can be considered the average value of $\dot{M}_{\rm in}$ during a time interval $\Delta t \sim t_{\rm SF} / ( 1 + \eta_M)$.

If $\dot{M}_{\rm in}$ and $t_{\rm SF}$ remain sufficiently steady, so that $\dot{M}_{\rm in} = \langle \dot{M}_{\rm in} \rangle$, then the minimalist regulator model reduces to
\begin{align}
    \dot{E}_{\rm CGM} \:
        &=  \: \dot{M}_{\rm acc} \varepsilon_{\rm acc}
                \: + \: \left( 
                    \frac {\eta_E \varepsilon_{\rm SN}} 
                            {1 + \eta_M}  
                    - \varepsilon_{\rm loss} \right)
                    \dot{M}_{\rm in}
        \label{eq:Edot_reduced}
            \\
    \dot{M}_{\rm CGM} \:
        &=  \: \dot{M}_{\rm acc}
                \: - \: \frac {\dot{M}_{\rm in}} {1 + \eta_M} 
            \; \; .
        \label{eq:Mdot_reduced}
\end{align}
We will use the minimalist regulator model in this reduced form whenever possible. Doing so enables an assessment of whether $\dot{M}_{\rm in}$ remains sufficiently steady, at least in the context of the model. If the steady-supply assumption is justified, then explicit integration of the $\dot{M}_{\rm ISM}$ equation is not necessary. The exceptions are cases with either an erratic gas supply or erratic variations in $t_{\rm SF}$. Explicit integration of the $\dot{M}_{\rm ISM}$ equation then smooths the star-formation response to the galaxy's gas supply over a time interval $\sim t_{\rm SF} / (1 + \eta_M)$.

\begin{figure}[t]
\begin{center}
  \includegraphics[width=0.49\textwidth]{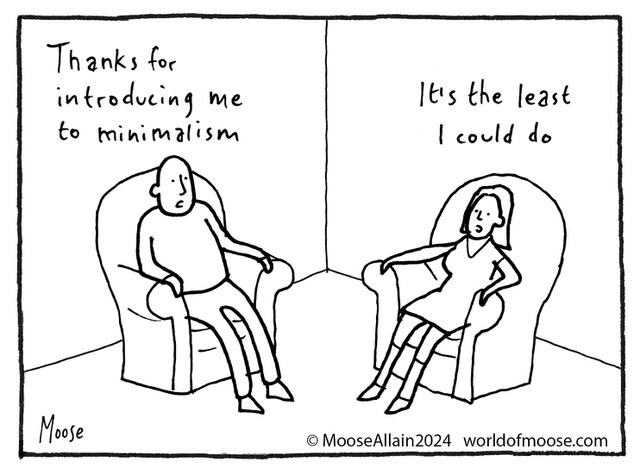}
\end{center}
\caption{Introduction to minimalism.
}
\label{fig:minimalism}
\end{figure}

\subsection{Generality of the Model}

Before moving on, we would like to emphasize the \textit{generality} of the minimalist regulator model. It is primarily an accounting system that tracks how baryons move from one reservoir to another and how cooling and feedback change the atmosphere's total energy. It becomes an integrable physical model once $\dot{M}_{\rm in}$ and $\varepsilon_{\rm loss}$ are specified as functions of $E_{\rm CGM}$, $M_{\rm CGM}$, and $t$. Those features make the model broadly useful for testing how various assumptions about the structure and radiative loss rate of a galaxy's atmosphere, as expressed via $\dot{M}_{\rm in}$ and $\varepsilon_{\rm loss}$, affect long-term predictions for galactic star formation. We will discuss $\dot{M}_{\rm in}$ and $\varepsilon_{\rm loss}$ further in \S \ref{sec:AtmosphereModels}.
 
\newpage

\section{Coupled Outflows: Asymptotic States}
\label{sec:Asymptotic}

Now we will use the minimalist regulator model to illuminate how mass loading of coupled galactic outflows affects the stellar baryon fraction $f_*$ of a galactic halo. At any given moment in time, there is a particular equilibrium state toward which the model is converging. That equilibrium state depends on the gas supply function $\dot{M}_{\rm in}$, the specific energy loss function $\varepsilon_{\rm loss}$, the cosmological baryon accretion rate $\dot{M}_{\rm acc}$, and the model's parameters: $\eta_M$, $\eta_E$, $\varepsilon_{\rm SN}$, and $\varepsilon_{\rm acc}$. In equilibrium, the star formation rate $\dot{M}_*$ and atmospheric specific energy $\varepsilon_{\rm CGM}$ remain constant. However, \textit{explicitly} time-dependent changes in either of the governing functions ($\dot{M}_{\rm in}$ or $\varepsilon_{\rm loss}$) or in the model's parameters can cause the equilibrium values of $\dot{M}_*$ and $\varepsilon_{\rm CGM}$ to drift with time.\footnote{To be clear, both $\dot{M}_{\rm in}$ and $\varepsilon_{\rm loss}$ may change with time because $E_{\rm CGM}$ and $M_{\rm CGM}$ are changing with time. Those temporal changes are \textit{implicit}. It is also possible for the dependences of those functions on $E_{\rm CGM}$ and $M_{\rm CGM}$ to change with time, and those kinds of temporal changes are \textit{explicit}.} In other words, there are potentially two types of time dependence: (1) asymptotic convergence toward an equilibrium state, and (2) temporal changes in the asymptotic equilibrium state.

The asymptotic states fall into two distinct categories, depending on the sign of the factor 
\begin{equation}
      \frac {\eta_E \varepsilon_{\rm SN}} 
                            {1 + \eta_M} 
            - \varepsilon_{\rm loss}
        \; \; .
\end{equation}
If this factor is positive, then the net effect of a coupled supernova feedback loop is to add energy to the surrounding atmosphere, causing the atmosphere to \textit{expand} and the central galaxy's gas supply to decline. If this factor is negative, then the atmosphere loses energy and \textit{contracts} relative to the state it would have without star formation and supernova feedback. 

This section will demonstrate why asymptotic star formation rates within \textit{expanded atmospheres} are insensitive to $\eta_M$ and will also explain why asymptotic star formation rates within \textit{contracted atmospheres} are likely to be larger than those within expanded atmospheres.

\subsection{Asymptotic Expansion}
\label{sec:Expansion}

Feedback that expands a galaxy's atmosphere causes star formation to approach a rate jointly determined by the cosmological accretion rate $\dot{M}_{\rm acc}$ and the specific energy $\eta_E \varepsilon_{\rm SN}$. For example, if $\varepsilon_{\rm loss}$ is negligible compared to $\eta_E \varepsilon_{\rm SN} / (1 + \eta_M)$, then essentially all of the feedback energy accumulates in the CGM, meaning that
\begin{equation}
    (\varepsilon_{\rm CGM} - \varepsilon_{\rm acc}) (\dot{M}_{\rm acc} - \dot{M}_*) 
        \: \approx \: \eta_E \varepsilon_{\rm SN} \dot{M}_*
\end{equation}
in which $\varepsilon_{\rm CGM} \equiv E_{\rm CGM} / M_{\rm CGM}$ is the mean specific energy of the CGM.  Significant expansion of the atmosphere generally requires a value of $\varepsilon_{\rm CGM} - \varepsilon_{\rm acc}$ similar to $v_{\rm c}^2$, and a value much greater than $v_{\rm c}^2$ would eject the entire atmosphere. Consequently, the halo's stellar baryon fraction asymptotically approaches
\begin{equation}
    f_* \: \approx \: 
            \frac {\xi v_{\rm c}^2} 
             {\eta_E \varepsilon_{\rm SN} 
                + \xi v_{\rm c}^2}
            \label{eq:fstar_xi}
\end{equation}
in which $\xi \equiv ( \varepsilon_{\rm eq} - \varepsilon_{\rm acc}) / v_{\rm c}^2$ is a dimensionless number of order unity and $\varepsilon_{\rm eq}$ is the value of $\varepsilon_{\rm CGM}$ that makes $\dot{\varepsilon}_{\rm CGM} = 0$. 

The absence of the mass loading parameter $\eta_M$ from this estimate for $f_*$ implies that the asymptotic stellar mass of the galaxy at the center of an expanding atmosphere is insensitive to mass loading, as long as radiative losses are negligible. Fundamentally, a supernova feedback system with these properties tunes itself so that the galaxy's \textit{energy} output is regulating its gas supply by keeping $\varepsilon_{\rm CGM}$ approximately constant and greater than $\varepsilon_{\rm acc}$ as cosmological accretion adds gas mass to the atmosphere. \citep[See also][who arrive at a similar conclusion through a different approach.]{SharmaTheuns_2020MNRAS.492.2418S}

\subsection{Asymptotic Contraction}
\label{sec:Contraction}

Radiative losses come into play when $\varepsilon_{\rm loss}$ is comparable to $\eta_E \varepsilon_{\rm SN} / (1 + \eta_M)$. If atmospheric energy losses exceed feedback energy input, then they reduce $E_{\rm CGM}$ below the cumulative energy input $E_{\rm acc}$ from cosmological accretion. Those losses eventually lead to an atmosphere with a smaller mean radius and greater mean density than an  atmosphere without radiative cooling or galaxy formation would have. Atmospheric contraction is therefore likely to increase the central galaxy's gas supply, because a denser atmosphere has both a shorter cooling time and a shorter dynamical time. 

Such an atmosphere is prone to runaway cooling and catastrophic collapse. The risk of collapse remains high while $\varepsilon_{\rm loss}$ remains greater than $\eta_E \varepsilon_{\rm SN} / (1 + \eta_M)$, because the net effect of additional star formation and supernova feedback is even more atmospheric energy loss. However, there are a few ways to prevent a catastrophe:
\begin{enumerate}

    \item Angular momentum can limit a galactic atmosphere's collapse once it contracts enough for orbital gas motions to prevent further contraction. The galaxy's gas supply then depends on torques that remove angular momentum from gas near the galaxy and on the timescale for dissipation of orbital energy.

    \item Accretion onto a central black hole can release additional feedback energy capable of exceeding radiative cooling and reversing the contraction.

    \item Atmospheric contraction itself can in principle reduce the specific energy loss function $\varepsilon_{\rm loss}$ so that it becomes comparable to $\eta_E \varepsilon_{\rm SN} / (1 + \eta_M)$. For example, consider gas in a supernova-driven galactic fountain that manages to keep most of a halo's baryons aloft in the CGM rather than allowing them to collect in the central galaxy. Gas in the fountain has a specific energy similar to $\varepsilon_{\rm fb} = \eta_E \varepsilon_{\rm SN} / \eta_M$ and therefore radiates an amount of energy comparable to $\varepsilon_{\rm fb}$ while returning to the galaxy.
    
\end{enumerate}
The first two remedies are beyond the scope of the minimalist regulator model, at least in the form presented in this paper. We will therefore focus here on the third possibility.

The threshold condition $\varepsilon_{\rm loss} = \eta_E \varepsilon_{\rm SN} / (1 + \eta_M)$ corresponds to a feedback energy input rate that balances atmospheric energy losses. Star formation in such a balanced state proceeds at a rate
\begin{equation}
    \dot{M}_* 
     \: = \: \frac {\dot{M}_{\rm in}} {1 + \eta_M}
     \: = \: \frac {\dot{E}_{\rm loss}} 
                   {\eta_E \varepsilon_{\rm SN}}
    \; \; .
\end{equation}
The atmosphere's energy loss rate $\dot{E}_{\rm loss}$ depends sensitively on the atmosphere's density and temperature structure. Consequently, the asymptotic value of $\dot{M}_*$ at the center of a contracted atmosphere also depends sensitively on atmospheric structure. 

We will not attempt to model here the complex details that determine $\dot{E}_{\rm loss}$ in a contracted atmosphere supported by supernova-driven outflows. Even sophisticated numerical simulations have trouble with that task. For the moment, we will simply point out two important global features of contracted atmospheres: (1) larger values of $\eta_M$ make an outflow's specific energy $\varepsilon_{\rm fb}$ smaller, resulting in atmospheres with smaller radii, and (2) denser atmospheres usually suffer greater radiative losses. Coupled outflows with larger values of $\eta_M$ are therefore likely to result in larger steady-state star formation rates. The next section tries to make that inference more quantitative.

\subsection{Analysis of Equilibria}
\label{sec:EquilibriumAnalysis}

Here we analyze how the minimalist regulator model for coupled outflows evolves toward an equilibrium state and how that asymptotic state depends on the model's parameters. Our goal is to understand how star formation depends on $\eta_M$ in such models.

\subsubsection{Dynamical Equations}

The following investigation of the model's equilibrium states focuses on the quantity
\begin{equation}
    f_{*,{\rm asy}} 
        \: \equiv \:
    \frac {\dot{M}_{\rm in}} {(1 + \eta_M) \dot{M}_{\rm acc}}
    \label{eq:fstar_asy}
    \; \; ,
\end{equation}
which is the asymptotic stellar baryon fraction of a halo with constant $\dot{M}_{\rm acc}$ and $\dot{M}_{\rm in}$. When written in terms of $f_{*,{\rm asy}}$, the mimimalist regulator model becomes
\begin{align}
    \frac {d E_{\rm CGM}} {d M_{\rm acc}} &= 
        \varepsilon_{\rm acc}
            \: + \: \left[ \eta_E \varepsilon_{\rm SN}
                    - (1 + \eta_M) \varepsilon_{\rm loss}
                    \right] f_{*,{\rm asy}} \nonumber \\
            &\quad\quad \: + \: \eta_E \varepsilon_{\rm SN}
                \cdot \delta f_* 
        \label{eq:dECGM_dMacc} \\
    \frac {d M_{\rm CGM}} {d M_{\rm acc}} &=
        1 \: - \: f_{*,{\rm asy}} + \eta_M \cdot \delta f_* 
        \label{eq:dMCGM_dMacc} \\[6 pt]
    \frac {d M_{\rm ISM}} {d M_{\rm acc}} &=
        - \, (1 + \eta_M) \cdot \delta f_*
        \label{eq:dMISM_dMacc}
        \; \; .
\end{align}
In these equations, the total mass of accreted baryons ($M_{\rm acc}$) plays the role of time and the quantity
\begin{equation}
    \delta f_* \: \equiv \: 
        \frac {\dot{M}_*} 
                {\dot{M}_{\rm acc}}
        \: - \: f_{*,{\rm asy}}
\end{equation}
accounts for differences between $\dot{M}_{\rm in} / (1 + \eta_M)$ and $\dot{M}_*$. Setting $\delta f_*$ equal to zero makes this system of equations equivalent to the reduced version of the minimalist regulator model defined by equations (\ref{eq:Edot_reduced}) and (\ref{eq:Mdot_reduced}).

\subsubsection{Equilibrium States}
\label{sec:Equilibria}

We would like to identify equilibrium states in which $\dot{M}_*$ remains constant as long as all of the model's parameters ($\varepsilon_{\rm acc}$, $\varepsilon_{\rm SN}$, $\eta_E$, and $\eta_M$) remain constant. In such an equilibrium state, $M_{\rm ISM} = \dot{M}_* t_{\rm SF}$ remains constant, implying $\delta f_* = 0$. The reduced version of the minimalist regulator model then applies. Also, the mass-exchange rates ($\dot{M}_*$, $\dot{M}_{\rm in}$, and $\dot{M}_{\rm acc}$) in such an equilibrium state must all remain proportional to each other as $E_{\rm CGM}$ and $M_{\rm CGM}$ increase with time. 

We will therefore restrict our attention to gas supply functions having the separable form
\begin{equation}
    \dot{M}_{\rm in} = f_{\rm CGM} \dot{M}_{\rm acc} \cdot
                        f_{\rm in} (\varepsilon_{\rm CGM}) 
    \label{eq:GasSupply}
\end{equation}
so that $\dot{M}_{\rm in}/\dot{M}_{\rm acc}$ does not change as long as the atmosphere's specific energy ($\varepsilon_{\rm CGM}$) and circumgalactic baryon fraction ($f_{\rm CGM} \equiv M_{\rm CGM} / M_{\rm acc}$) remain constant. The quantity $f_{\rm CGM} \dot{M}_{\rm acc} = M_{\rm CGM} / (M_{\rm acc} / \dot{M}_{\rm acc})$ corresponds to the gas supply rate provided by an atmosphere of mass $M_{\rm CGM}$ that flows into a halo's central galaxy on the cosmological timescale $M_{\rm acc} / \dot{M}_{\rm acc}$. The dimensionless function $f_{\rm in} (\varepsilon_{\rm CGM})$ accounts for how cooling and feedback change the gas supply rate by causing the atmosphere either to expand or to contract. 

When expressed in terms of $\varepsilon_{\rm CGM}$ and $f_{\rm CGM}$, the reduced version of the minimalist regulator model becomes
\begin{align}
    \frac {d \varepsilon_{\rm CGM}} {d \ln M_{\rm acc}} &= 
        \varepsilon_{\rm acc}
            \: + \: \left[ \eta_E \varepsilon_{\rm SN}
                    - (1 + \eta_M) \varepsilon_{\rm loss}
                    \right] f_{*,{\rm asy}} \nonumber \\
            &\quad\quad \: - \: ( 1 - f_{*,{\rm asy}}) \varepsilon_{\rm CGM} 
        \label{eq:deps_dMacc} \\
    \frac {d f_{\rm CGM}} {d \ln M_{\rm acc}} &=
        1 \: - \: f_{*,{\rm asy}} - f_{\rm CGM} 
        \label{eq:dfCGM_dMacc} 
        \; \; .
\end{align}
A gas supply function with the form of equation (\ref{eq:GasSupply}) remains constant if $\varepsilon_{\rm CGM}$, $f_{\rm CGM}$, and $\dot{M}_{\rm acc}$ all remain constant. The equilibrium values of $\varepsilon_{\rm CGM}$ and $f_{\rm CGM}$ are then solutions to the system of equations
\begin{eqnarray}
    \varepsilon_{\rm CGM}
        & \: = \: &  \frac 
            {\varepsilon_{\rm acc}
            \: + \: \left[ \eta_E \varepsilon_{\rm SN}
                    - (1 + \eta_M) \varepsilon_{\rm loss}
                    \right] f_{*,{\rm asy}}}
            {1 - f_{*,{\rm asy}}}
        \label{eq:eps_eq}
        \\[4pt]
    f_{\rm CGM} 
        &  \: = \:  & 1 \: - \: f_{*,{\rm asy}} 
        \label{eq:fCGM_eq}
\end{eqnarray}
in which $f_{*,{\rm asy}}$ and $\varepsilon_{\rm loss}$ are functions of $\varepsilon_{\rm CGM}$ and $f_{\rm CGM}$. 

Combining equations (\ref{eq:fstar_asy}), (\ref{eq:GasSupply}), and (\ref{eq:fCGM_eq}) gives the asymptotic stellar baryon fraction
\begin{equation}
    f_{*,{\rm asy}}(\varepsilon_{\rm eq}) 
        \: = \: \frac {f_{\rm in} (\varepsilon_{\rm eq})} 
                            {1 + \eta_M + f_{\rm in} (\varepsilon_{\rm eq})}
    \label{eq:fstar_eq1}
\end{equation}
in which $\varepsilon_{\rm eq}$ is the equilibrium value of $\varepsilon_{\rm CGM}$. It is also possible to obtain
\begin{equation}
    f_{*,{\rm asy}}(\varepsilon_{\rm eq})
        \: = \: \frac {\varepsilon_{\rm eq} 
                                    - \varepsilon_{\rm acc}}
                          {\eta_E \varepsilon_{\rm SN} 
                           + \varepsilon_{\rm eq}
                           - (1 + \eta_M) \varepsilon_{\rm loss} (\varepsilon_{\rm eq})}
            \label{eq:fstar_eq2}    
\end{equation}
by rearranging equation (\ref{eq:eps_eq}). The latter equation expresses $\varepsilon_{\rm loss}$ as a function of just $\varepsilon_{\rm eq}$, because the equilibrium value of $f_{\rm CGM}$ depends only on $\varepsilon_{\rm eq}$, via equations (\ref{eq:fCGM_eq}) and (\ref{eq:fstar_eq1}).


\newpage

\subsubsection{Astrophysical Interpretation}
\label{sec:Interpretation}

Before proceeding with the analysis, it's worth spending a moment considering the astrophysical implications of equations (\ref{eq:fstar_eq1}) and (\ref{eq:fstar_eq2}). Notice that equation (\ref{eq:fstar_eq1}) reflects the amount of recycling implicit in the minimalist regulator model. For example, if $f_{\rm in}(\varepsilon_{\rm eq})$ is much less than unity, then only a small fraction of a halo's gas has cycled through its central galaxy, and the proportion that has formed stars is $f_{\rm in}(\varepsilon_{\rm eq})/(1 + \eta_M)$. Conversely, if $f_{\rm in}(\varepsilon_{\rm eq})$ is much greater than unity, then halo gas typically cycles through the central galaxy many times, and a fraction $1/(1 + \eta_M)$ forms stars during each cycle. Consequently, a halo's stellar baryon fraction approaches unity when $f_{\rm in}(\varepsilon_{\rm eq})$ exceeds $1 + \eta_M$.


Equation (\ref{eq:fstar_eq2}) is a quantitative and succinct representation of the physical reasoning in \S \ref{sec:Expansion} and \S \ref{sec:Contraction}. Rewriting it terms of $\xi \equiv (\varepsilon_{\rm eq} - \varepsilon_{\rm acc})/v_{\rm c}^2$ makes its relationship to equation (\ref{eq:fstar_xi}) more transparent:
\begin{equation}
    f_{*,{\rm asy}} =  \frac {\xi v_{\rm c}^2} 
        {\eta_E \varepsilon_{\rm SN}+ \xi v_{\rm c}^2 
            + [\varepsilon_{\rm acc} - (1 + \eta_M) \varepsilon_{\rm loss}] }
            \label{eq:fstar_eq3}
            \; \; .
\end{equation}
The term in square brackets accounts for radiative losses, meaning that equation (\ref{eq:fstar_eq2}) reduces to equation (\ref{eq:fstar_xi}) when that term is small.  In that limit, supernova feedback acts to maintain the expanded state of a galaxy's atmosphere as cosmological accretion proceeds. The galaxy's equilibrium star formation rate then depends more directly on $\eta_E$ than on $\eta_M$. However, the nature of the atmosphere's equilibrium state qualitatively changes as radiative energy losses become comparable to supernova energy input, for the reasons discussed in \S \ref{sec:Contraction}.

According to equation (\ref{eq:fstar_eq2}), the qualitative transition happens as the system passes through the singular case with $\varepsilon_{\rm eq} = \varepsilon_{\rm acc}$, which corresponds to a critical value of the mass-loading parameter:
\begin{equation}
    \eta_M = \frac {\eta_E \varepsilon_{\rm SN} 
                                + \varepsilon_{\rm acc}
                                - \varepsilon_{\rm loss} 
                                    (\varepsilon_{\rm acc})}
                            {\varepsilon_{\rm loss} 
                                    (\varepsilon_{\rm acc})}
            \; \; .
\end{equation}
When mass loading of coupled outflows exceeds this critical value, then a galaxy's atmosphere converges toward an equilibrium state with \textit{less} specific energy than the specific energy of accreting gas. Supernova heating must somehow balance atmospheric radiative losses in that state. Furthermore, if we assume that the specific energy losses necessary for gas to enter a galaxy are similar to the specific energy of accreting gas, so that $\varepsilon_{\rm loss} (\varepsilon_{\rm acc}) \approx \varepsilon_{\rm acc}$, we then obtain the approximation
\begin{equation}
    \eta_M
        \approx \frac {\eta_E \varepsilon_{\rm SN}} 
                      {\varepsilon_{\rm acc}}
\end{equation}
corresponding to the vertical line in Figure \ref{fig:FlipSideGraphic}.

\subsection{Atmosphere Models}
\label{sec:AtmosphereModels}

So far, we have avoided proposing a physical model for the gas supply function $\dot{M}_{\rm in}$ because the astrophysical details can be forbiddingly complex. In preparation for implementing such a physical model, we have constructed and analyzed an accounting framework---the minimalist regulator model---that sorts a halo's baryonic mass and energy into categories and tracks how mass and energy move from one reservoir to another. While doing that, we have made few astrophysical assumptions so that the framework broadly applies to galactic baryon cycles dominated by supernova feedback.\footnote{Adding a feedback term accounting for black-hole energy injection would make it even more general.}

Within this framework, all of the fearsome astrophysical details end up bundled into the gas supply function $\dot{M}_{\rm in}$ and the energy-loss function $\varepsilon_{\rm loss}$. Together, those two functions represent an \textit{atmosphere model} describing a particular astrophysical scenario, in which the product $\dot{M}_{\rm in} \varepsilon_{\rm loss}$ is equal to the sum of the atmosphere's radiative loss rate $\dot{E}_{\rm rad}$ and the rate $\dot{E}_{\rm in}$ at which gas flowing into the halo's central galaxy transports energy out of the CGM (see \S \ref{sec:Dynamics}). Both $\dot{M}_{\rm in}$ and $\varepsilon_{\rm loss}$ can be specified as functions of $E_{\rm CGM}$, $M_{\rm CGM}$, and $t$, or equivalently, as functions of $\varepsilon_{\rm CGM}$, $f_{\rm CGM}$, and $t$. Differing astrophysical scenarios\footnote{These could include cooling flows, cold streams, precipitation models, or galactic fountains, to name a few examples.} can then be tested by inserting various choices for the $\dot{M}_{\rm in}$ and $\varepsilon_{\rm loss}$ functions into the minimalist regulator model and integrating it over a particular gas-mass accretion history $\dot{M}_{\rm acc}(t)$.

Future papers in this series will present specific families of physically motivated atmosphere models and will analyze how they evolve. Here we will illustrate the general procedure by examining the properties of a generic atmosphere model that provides some useful insights.

\subsubsection{Generic Atmosphere Model}
\label{sec:GenericModel}

To build intuition about the dependence of a galactic atmosphere's equilibrium states on $\eta_M$ without having to construct a more complicated model, we will examine the equilibria and phase plane trajectories of a toy atmosphere model defined by $\varepsilon_{\rm loss} = \varepsilon_{\rm CGM}$ and 
\begin{equation}
    f_{\rm in} (\varepsilon_{\rm CGM}) 
            = \exp \left( \frac {\varepsilon_{\rm acc} 
                                 - \varepsilon_{\rm CGM}}
                                   {v_{\rm c}^2} \right)
\end{equation}
Our motivation for making $f_{\rm in}$ an exponential function of $\varepsilon_{\rm CGM} / v_{\rm c}^2$ is that the characteristic radius and dynamical time of an atmosphere in an isothermal potential well are both proportional to $\exp(\varepsilon_{\rm CGM}/v_{\rm c}^2)$. 

The normalization of the exponential function corresponds to a gas supply rate that would transport gas into the halo's central galaxy at a rate $\dot{M}_{\rm in} \approx \dot{M}_{\rm acc}$ as long as $\varepsilon_{\rm CGM} \approx \varepsilon_{\rm acc}$ and $f_{\rm CGM} \approx 1$. When feedback heating exceeds radiative cooling, the gas supply exponentially declines as $\varepsilon_{\rm CGM}$ rises. Conversely, when radiative cooling exceeds feedback heating, the gas supply exponentially rises as $\varepsilon_{\rm CGM}$ falls.

\subsubsection{The Missing Details}

Many readers who have reached this point in the paper may be wondering why they have not yet seen a radiative cooling function, an assessment of heavy-element enrichment, a description of atmospheric structure, or a breakdown of multiphase atmospheric components and their interactions. We have not needed to specify those details because the equilibrium states of the minimalist regulator model can be identified and classified without them. Also, the asymptotic stellar baryon fraction of a halo with an expanded atmosphere does not depend much on the details, because equilibrium states with $\varepsilon_{\rm fb} > \varepsilon_{\rm acc}$ are not sensitive to radiative cooling.

The missing details become more important in contracted atmospheres, because the gas supply rate $\dot{M}_{\rm in}(\varepsilon_{\rm eq})$ associated with the atmosphere's equilibrium state is then closely tied to its radiative energy loss rate. However, galactic atmospheres in the most recent generation of cosmological numerical simulations tend to be expanded, not contracted, with closure radii that can be several times the halo's virial radius \citep[e.g.,][]{Ayromlou_2023MNRAS.524.5391A,Sorini_2022MNRAS.516..883S,Wright2024}. It therefore appears likely that long-term galactic star formation rates depend mostly on the self-regulating properties of atmospheric expansion, as described in \S {\ref{sec:Expansion}.

To clarify, radiative cooling is still essential for star formation in a galaxy at the center of an expanded atmosphere. Its gas supply and long-term star formation rates are still proportional to the atmosphere's radiative cooling rate. But its \textit{equilibrium state} depends on the interplay between supernova feedback and cosmological accretion, not radiative cooling.

\subsection{Sensitivity to $\eta_M$}
\label{sec:MassLoadingSensitivity}

\begin{figure}[t]
\begin{center}
  \includegraphics[width=0.49\textwidth]{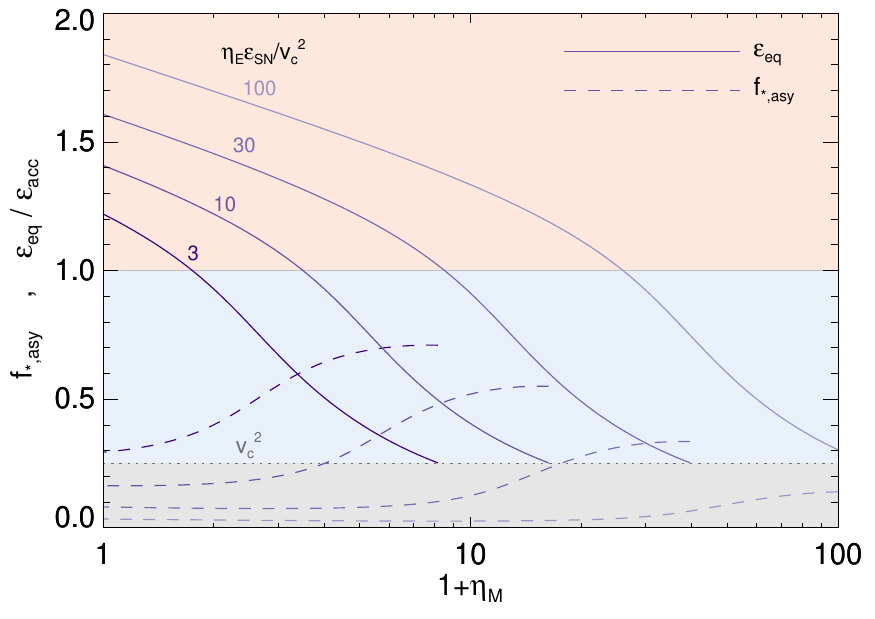}
\end{center}
\caption{Dependence of equilibrium specific energy $\varepsilon_{\rm eq}$ (solid lines) and asymptotic stellar baryon fraction $f_{*,{\rm asy}}$ (dashed lines) on the mass loading parameter $\eta_M$ in the generic atmosphere model of \S \ref{sec:GenericModel}. Colors represent supernova feedback strength $\eta_E \varepsilon_{\rm SN} / v_{\rm c}^2$ as labeled. A horizontal solid line marks $\varepsilon_{\rm eq} = \varepsilon_{\rm acc}$. A horizontal dotted line marks $\varepsilon_{\rm eq} = v_{\rm c}^2$. All examples have $\varepsilon_{\rm acc} = 4 v_{\rm c}^2$. Atmospheres with $\varepsilon_{\rm eq}$ in the red shaded area have expanded. Atmospheres with $\varepsilon_{\rm eq}$ in the blue shaded area have contracted. Each line is truncated as $\varepsilon_{\rm eq}$ drops into the grey region because the generic atmosphere model breaks at small values of $\varepsilon_{\rm CGM}$.
}
\label{fig:fstar_asy_multi_plot}
\end{figure}

Now we are ready to consider the role of mass loaded galactic winds.
Figure \ref{fig:fstar_asy_multi_plot} illustrates how the generic atmosphere's equilibrium specific energy ($\varepsilon_{\rm eq}$) and asymptotic stellar baryon fraction ($f_{*,{\rm asy}}$) depend on $\eta_M$. Four color-coded pairs of lines show their equilibrium values for a particular choice of supernova feedback strength, quantified by $\eta_E \varepsilon_{\rm SN} / v_{\rm c}^2$. Solid lines represent $\varepsilon_{\rm eq}/\varepsilon_{\rm acc}$, dashed lines represent $f_{*,{\rm asy}}$, and all examples have $\varepsilon_{\rm acc} = 4 v_{\rm c}^2$.

As expected, halos with expanded atmospheres (i.e. $\varepsilon_{\rm eq} > \varepsilon_{\rm acc}$) have stellar baryon fractions that are insensitive to mass loading. When an atmosphere's specific energy (solid line) is in the red region of the figure, the corresponding dashed line representing $f_{*,{\rm asy}}$ remains nearly flat as $\eta_M$ increases.

Also as expected, each of the solid lines falls below $\varepsilon_{\rm eq} = \varepsilon_{\rm acc}$ and crosses into the blue (contracted) region of the figure as $\eta_M$ surpasses $\eta_E \varepsilon_{\rm SN} / \varepsilon_{\rm acc}$.  When that happens, the corresponding value of $f_{*,{\rm asy}}$ becomes more sensitive to the mass-loading parameter and sharply rises as $\eta_M$ increases. 

Each line is then truncated where $\varepsilon_{\rm eq}$ drops into the grey region below $\varepsilon_{\rm eq} = v_{\rm c}^2$. That is where $\dot{M}_{\rm in}$ becomes insensitive to $\varepsilon_{\rm CGM}$, according to the generic atmosphere model. Qualitatively, the atmosphere enters a regime in which its altitude is similar to the galaxy's size. Ejected gas then falls back into the galaxy on a timescale similar to the galaxy's dynamical time. A different approach for estimating $\dot{M}_{\rm in}$ is needed in this limit because the CGM and ISM are becoming indistinguishable (see also \S \ref{sec:Circulation}).

Some readers may be surprised by how star formation rises as $\eta_M$ increases beyond $\eta_E \varepsilon_{\rm SN} / \varepsilon_{\rm acc}$, a trend opposite to the usual intuitive assumptions about mass loading. It comes about because greater mass loading factors result in outflows that fail to replace the losses of atmospheric specific energy that lead to star formation (i.e. $\varepsilon_{\rm fb} < \varepsilon_{\rm acc}$). When an outflow's specific energy is less than the atmosphere's mean specific energy, it reduces the atmosphere's mean specific energy and promotes atmospheric contraction, allowing recycling to be more rapid.

Recycling implied by the generic atmosphere model is exponentially sensitive to $\varepsilon_{\rm CGM}$ because the recycling rate is proportional to $f_{\rm in}(\varepsilon_{\rm CGM})$. Consequently, recycling is also exponentially sensitive to the effects that $\eta_M$ has on  $\varepsilon_{\rm CGM}$. As a result, the galaxy's gas supply $\dot{M}_{\rm in}$ increases more quickly than $1 + \eta_M$ as $\eta_M$ increases, and so the steady-state star formation rate $\dot{M}_* = \dot{M}_{\rm in} / (1 + \eta_M)$ also increases.

Many plausible atmosphere models share this qualitative property with the generic atmosphere model. The key is exponential sensitivity of the gas supply to $\eta_M$. It arises in the generic atmosphere model because the timescale for recycling scales with the atmosphere's mean radius, which is exponentially sensitive to $\varepsilon_{\rm CGM}$. It would also arise in a cooling-flow model that feeds gas into the central galaxy on a cooling time scale, because cooling time is inversely proportional to density, and the mean density of a galactic atmosphere is also exponentially sensitive to $\varepsilon_{\rm CGM}$. Future papers investigating more sophisticated atmosphere models will explore these relationships in greater detail.

\subsection{Convergence toward Equilibrium}

\begin{figure*}[t]
\begin{center}
  \includegraphics[width=0.9\textwidth]{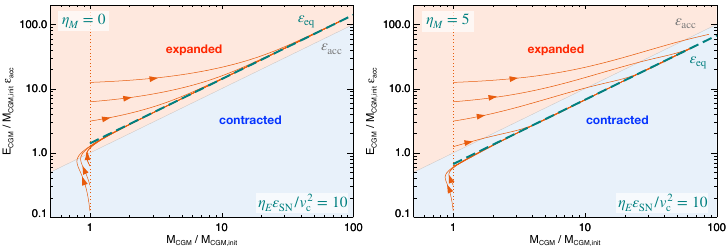}
\end{center}
\caption{Trajectories of generic atmosphere models in the $M_{\rm CGM}$--$E_{\rm CGM}$ plane. Both panels show models with $\varepsilon_{\rm acc}/v_{\rm c}^2 = 4$ and $\eta_E \varepsilon_{\rm SN}/v_{\rm c}^2 = 10$. Trajectories in the left panel (solid orange lines) have $\eta_M = 0$, and trajectories in the right panel have $\eta_M = 5$. They all start at $M_{\rm CGM} = M_{\rm CGM,init}$ with various values of $E_{\rm CGM}$ and converge toward the dashed green lines along which $E_{\rm CGM} / M_{\rm CGM} = \varepsilon_{\rm eq}$. Grey lines indicate $\varepsilon_{\rm CGM} = \varepsilon_{\rm acc}$. Models with less mass loading end up in the red (expanded) region with $\varepsilon_{\rm CGM} > \varepsilon_{\rm acc}$ Models more mass loading end up in the blue (contracted) region with $\varepsilon_{\rm CGM} < \varepsilon_{\rm acc}$. 
}
\label{fig:MCGM_ECGM_trajectories}
\end{figure*}

\begin{figure*}
\begin{center}
  \includegraphics[width=0.9\textwidth]{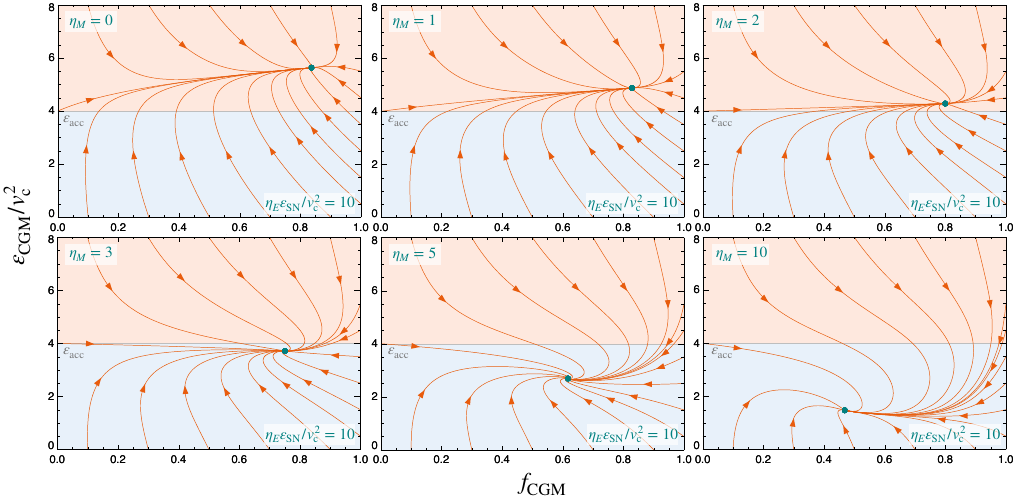}
\end{center}
\caption{Trajectories of generic atmosphere models in the $f_{\rm CGM}$--$\varepsilon_{\rm CGM}$ phase plane. All examples have $\varepsilon_{\rm acc}/v_{\rm c}^2 = 4$ and $\eta_E \varepsilon_{\rm SN}/v_{\rm c}^2 = 10$. The panels show trajectories for $\eta_M = 0$, 1, 2, 3, 5, and 10, as labeled. Those trajectories (orange lines) converge toward the green dots marking each atmosphere's equilibrium configuration. Excess heating leads to equilibria in the red (expanded) region ($\varepsilon_{\rm CGM} > \varepsilon_{\rm acc}$). Excess cooling leads to equilibria in the blue (contracted) region ($\varepsilon_{\rm CGM} < \varepsilon_{\rm acc}$).
}
\label{fig:phaseplane_plots_10}
\end{figure*}

We will conclude \S \ref{sec:Asymptotic} with a look at the convergence properties of the minimalist regulator model. Figure \ref{fig:MCGM_ECGM_trajectories} shows model trajectories in the $M_{\rm CGM}$--$E_{\rm CGM}$ plane for two different values of $\eta_M$. Each trajectory comes from integrating a reduced version of the minimalist regulator model that incorporates the generic atmosphere model of \S \ref{sec:GenericModel}. The integrations begin with various values of $E_{\rm CGM}$ and a circumgalactic medium comprising all of the halo's baryons, so that $f_{\rm CGM} = 1$. All models have $\varepsilon_{\rm acc} / v_{\rm c}^2 = 4$ and $\eta_E \varepsilon_{\rm SN}/v_{\rm c}^2 = 10$. The dimensionless quantity $\eta_E \varepsilon_{\rm SN}/v_{\rm c}^2$ determines how effectively supernova feedback lifts gas within the halo's potential well. The chosen value of that quantity corresponds to a galaxy with $\eta_E \approx 1$ and a potential well similar to the Milky Way's. 

Models with negligible mass loading ($\eta_M = 0$) converge toward an equilibrium value of atmospheric specific energy ($\varepsilon_{\rm eq}$) exceeding the specific energy of accreting gas ($\varepsilon_{\rm acc}$) because cumulative supernova heating of the atmosphere exceeds cumulative radiative cooling. Those models consequently result in expanded atmospheres. In contrast, the models with substantial mass loading ($\eta_M = 5$) converge toward an equilibrium specific energy smaller than $\varepsilon_{\rm acc}$, because cumulative radiative cooling exceeds cumulative supernova heating, resulting in a contracted atmosphere.

How an atmosphere model's convergence toward equilibrium depends on $\eta_M$ is easier to appreciate in a phase plane with $f_{\rm CGM}$ on one axis and $\varepsilon_{\rm CGM}$ on the other. An equilibrium state of the minimalist regulator model is then a fixed point rather than a line. Figure \ref{fig:phaseplane_plots_10} illustrates six examples based on the generic atmosphere model of \S \ref{sec:GenericModel}. All of them have $\varepsilon_{\rm acc} / v_{\rm c}^2 = 4$ and $\eta_E \varepsilon_{\rm SN}/v_{\rm c}^2 = 10$ but differing values of the mass loading parameter $\eta_M$. As in Figure \ref{fig:MCGM_ECGM_trajectories}, each orange line represents an integration of a reduced version of the minimalist regulator model incorporating the generic atmosphere model.

A green dot in each panel marks the model's equilibrium point, at which $\varepsilon_{\rm CGM}$ and $f_{\rm CGM}$ remain constant, as long as the halo's accretion rate remains constant. Far from equilibrium, each example evolves more rapidly in the vertical direction than in the horizontal direction, meaning that either net heating or net cooling causes the atmosphere's specific energy to change more rapidly than the ratio of galactic baryons to circumgalactic baryons. Once an atmosphere model is near its equilibrium value of $\varepsilon_{\rm CGM}$, it converges horizontally toward its equilibrium value of $f_{\rm CGM}$.

Mass loading is relatively modest in the upper row of examples, with $\eta_M = 0$, 1, and 2. Those examples have equilibrium points in the red (expanded) region of each panel, with $\varepsilon_{\rm CGM} > \varepsilon_{\rm acc}$, meaning that cumulative feedback heating exceeds cumulative radiative cooling. In those examples the equilibrium value of the asymptotic stellar baryon fraction ($f_{*,{\rm asy}} = 1 - f_{\rm CGM} \approx 0.2$) is insensitive to the mass loading parameter.

However, further increases of $\eta_M$ reduce the proportion of accreted baryons that remain circumgalactic and therefore increase the halo's asymptotic stellar baryon fraction. As $\eta_M$ increases from 2 to 10, the atmospheric equilibrium point drops into the blue (contracted) region with $\varepsilon_{\rm CGM} < \varepsilon_{\rm acc}$, meaning that cumulative radiative cooling exceeds cumulative feedback heating. The asymptotic stellar baryon fraction of the equilibrium state simultaneously rises, exceeding 50\% in the example with $\eta_M = 10$.

\section{Uncoupled Outflows and Circulation}
\label{sec:Ballistic}

All of the predictions discussed in \S \ref{sec:Dynamics} and \S \ref{sec:Asymptotic} pertain to \textit{coupled} supernova-driven outflows that deposit energy into the circumgalactic medium. For completeness, this section considers how mass loaded outflows affect star formation if they are \textit{not} coupled with the rest of a galaxy's atmosphere and therefore \textit{cannot} reduce the galaxy's gas supply. In particular, it focuses on why uncoupled outflows with large mass-loading factors can result in excessive recycling.

\subsection{Recycling and Star Formation Rate}

When recycling of a galaxy's atmosphere happens, the relationship between $\dot{M}_*$ and $\eta_M$ depends on the number of times baryons typically cycle through the galaxy ($N_{\rm cyc}$). To make this statement more quantitative, let us suppose that $\dot{M}_{\rm in,0}$ is the rate at which halo gas enters a galaxy for the first time, fueling star formation that ejects a proportion $\eta_M/(1 + \eta_M)$ of that gas. If the ejected gas eventually falls back into the galaxy, its infall rate has been reduced by the factor $\eta_M/(1 + \eta_M)$. After $n$ cycles, it has been reduced by the factor $[ \eta_M/(1 + \eta_M) ]^n$. The total gas infall rate at time $t$ in this idealized recycling model is therefore
\begin{equation}
    \dot{M}_{\rm in} (t) = \sum_{n = 0}^{N_{\rm cyc}} 
                            \left( \frac {\eta_M} {1 + \eta_M} \right)^n 
                            \dot{M}_{\rm in,0}(t - n t_{\rm cyc})
                            \; \; ,
\end{equation}
where $t_{\rm cyc} = t / N_{\rm cyc}$ is the time to complete one cycle.

Now suppose the initial gas supply rate $\dot{M}_{\rm in,0}$ is approximately constant with time, so it can be taken out of the sum, leaving the polynomial $1 + x + ... + x^{N_{\rm cyc}}$, where $x = \eta_M / (1 + \eta_M)$. In that case, the sum reduces to $(1 - x^{N_{\rm cyc}+1})/(1 - x)$, leading to
\begin{equation}
    \dot{M}_{\rm in} \approx (1 + \eta_M) 
                             \left[ 1 - \left( \frac {\eta_M} {1 + \eta_M} \right)^{N_{\rm cyc}+1} \right] 
                             \dot{M}_{\rm in,0}
                            \; \; .
\end{equation}
Dividing by $1+\eta_M$ then gives the steady-state star formation rate that the galaxy asymptotically approaches:
\begin{equation}
    \dot{M}_* \: \approx \: \left[ 1 - \left( \frac {\eta_M} {1 + \eta_M} \right)^{N_{\rm cyc}+1} \right] 
                             \dot{M}_{\rm in,0}
                            \; \; .
\end{equation}
Notice that $\dot{M}_*$ converges toward $\dot{M}_{\rm in,0}$ when $N_{\rm cyc}$ is sufficiently large, because excessive recycling ends up turning a galaxy's \textit{entire} gas supply into stars, if given enough time.

This expression can be further simplified when $N_{\rm cyc} \ll \eta_M$. The sum then reduces to $\dot{M}_{\rm in} \approx (N_{\rm cyc} + 1) \dot{M}_{\rm in,0}$, giving
\begin{equation}
    \frac {\dot{M}_*} {\dot{M}_{{\rm in},0}}  
        \: \approx \: \frac {N_{\rm cyc}+1} {1 + \eta_M}
        \label{eq:RecyclingSFR}
      \; \; ,
\end{equation}
because each of the $N_{\rm cyc} + 1$ terms in the sum is close to unity. This result supports the more intuitive argument presented in \S \ref{sec:Foundation} and may seem to support the simple picture of decreasing star formation by increasing mass loading. However, we still need to account for the hidden dependence of $N_{\rm cyc}$ on $\eta_M$. The next section presents an estimate for that relation.

\subsection{Ballistic Circulation}
\label{sec:Circulation}

Here we will consider uncoupled outflows that undergo what we will call \textit{ballistic circulation}. During ballistic circulation, a clump of ejected gas transfers little to none of its ejection energy to the rest of the atmosphere. Each clump then follows its own nearly ballistic trajectory, reaching a maximum altitude determined by the specific energy 
\begin{equation}
    \varepsilon_{\rm fb} = \frac {\eta_E \varepsilon_{\rm SN}} {\eta_M}    
\end{equation}
it had upon leaving the galaxy. Given enough time, those clumps fall back toward the galaxy and augment its gas supply, potentially entering and leaving the galaxy several times if the outflows are significantly mass loaded.\footnote{Some astronomers call this feedback mode ``juggling."}  

Ejected gas cannot formally escape the isothermal potential well we have adopted, because $\varphi (r)$ logarithmically approaches infinity as $r$ increases. However, gas ejected with a specific energy much greater than the gas currently accreting onto the halo does not fall back down on a timescale short enough to be recycled, for reasons discussed in \S \ref{sec:Foundation}. Consequently, ballistic gas in outflows with 
\begin{equation}
    \eta_M > \frac {\eta_E \varepsilon_{\rm SN}}        
                    {\varepsilon_{\rm acc}}
\end{equation}
cycles back into a halo's central galaxy, and ballistic outflows with less mass loading decouple from the central galaxy's gas supply.

In a nearly isothermal potential well, the timescale $t_{\rm cyc}$ for ballistic recycling is approximately proportional to the maximum radius of a ballistic trajectory with specific energy $\varepsilon_{\rm fb}$, which is $\sim r_0 \exp (\varepsilon_{\rm fb}/{v_{\rm c}^2})$. We therefore conclude that
\begin{equation}
    N_{\rm cyc} = \frac {t} {t_{\rm cyc}}
        \sim \exp \left( \frac {\varepsilon_{\rm acc} -  \varepsilon_{\rm fb}}
                            {v_{\rm c}^2} \right)
        \label{eq:BallisticNcyc}
\end{equation}
for recycling within a nearly isothermal halo.

With this approximation for $N_{\rm cyc}$, we can take the derivative of equation (\ref{eq:RecyclingSFR}) with respect to $\eta_M$, finding
\begin{equation}
    \frac {d } {d \eta_M} \frac {\dot{M}_*} {\dot{M}_{\rm in,0}} 
        \: \sim  \: \frac {1} {1 + \eta_M} 
                        \left( \frac {N_{\rm cyc}} {\eta_M} \frac {\varepsilon_{\rm fb}} {v_{\rm c}^2} 
                        - \frac {N_{\rm cyc}+1} {1 + \eta_M} \right)
                        \; \; .
\end{equation}
The derivative is negative, implying that increased mass loading lowers the galaxy's star formation rate, as long as
\begin{equation}
    N_{\rm cyc} \lesssim \left( \frac {1 + \eta_M} {\eta_M} \frac {\varepsilon_{\rm fb}} {v_{\rm c}^2} 
                            -1 \right)^{-1}
            \; \; .
\end{equation}  
Using equation (\ref{eq:BallisticNcyc}), we see that the derivative changes sign near where
\begin{eqnarray}
    \varepsilon_{\rm fb} \approx \varepsilon_{\rm acc} +
        v_{\rm c}^2 \ln \left( \frac {1 
         + \eta_M} {\eta_M} \frac {\varepsilon_{\rm fb}} {v_{\rm c}^2} - 1 \right) 
         \; \; .
\end{eqnarray}
Below this critical value of specific feedback energy, further increases of $\eta_M$ cause the galaxy's steady-state star formation rate to \textit{increase}. That happens because the galaxy's gas supply rate, enhanced by recycling, increases more rapidly than $1 + \eta_M$.

According to this approximation, recycling-driven enhancement of the galaxy's gas supply saturates as $\varepsilon_{\rm fb}$ approaches $v_{\rm c}^2$. But heavily mass-loaded galactic outflows don't propagate very far in that limit. They reach a maximum altitude $\lesssim 3 r_0$ and return to the galaxy on a timescale similar to the galaxy's rotation period. Such outflows are actually low-altitude galactic fountains. If most of the gas accreting onto a halo enters its central galaxy, perhaps through cold cosmological streams, then almost all of it ends up circulating at radii similar to the galaxy's radius.

Such a \textit{recycling crisis} would make the predicted star formation rate several times greater than the minimum value reached at $\eta_M \approx \eta_E \varepsilon_{\rm SN} / \varepsilon_{\rm acc}$ (see \S \ref{sec:Comparison}). In a halo suffering from a recycling crisis, the majority of the halo's baryons would be concentrated near its central galaxy. The rarity of low-redshift galaxies observed to have that property therefore indicates that ballistic circulation in the present-day universe, if it happens, results from outflows with $\eta_M < \eta_E \varepsilon_{\rm SN} / v_{\rm c}^2$. 

\subsection{Comparison with Coupled Outflows}
\label{sec:Comparison}

}

This schematic analysis of uncoupled outflows illustrates why excessive mass loading results in a crisis \textit{regardless} of whether supernova-driven outflows from galaxies are coupled or uncoupled. In both cases, feedback with $\eta_M > \eta_E \varepsilon_{\rm SN} / \varepsilon_{\rm acc}$ fails to push a halo's baryons beyond its virial radius. A crisis becomes inevitable because galactic gas returning to the circumgalactic medium has less specific energy than it had when it accreted onto the halo. Under those circumstances, the net effect of radiative cooling, star formation, and supernova feedback is a \textit{reduction} in atmospheric energy that allows the atmosphere to \textit{contract.} If no other form of feedback intervenes, then the resulting increase in atmospheric density and reduction in atmospheric dynamical time combine to boost the central galaxy's gas supply and star formation rate. Larger values of $\eta_M$ just make the crisis worse. That is why both curves in the bottom panel of Figure \ref{fig:FlipSideGraphic} rise toward larger $f_*$ as $\eta_M$ rises beyond $\eta_E \varepsilon_{\rm SN} / \varepsilon_{\rm acc}$

However, those curves diverge as $\eta_M$ drops below $\eta_E \varepsilon_{\rm SN} / \varepsilon_{\rm acc}$. Uncoupled outflows become \textit{less} effective at suppressing star formation because feedback energy becomes concentrated in a smaller proportion of the halo's baryons and fails to suppress the galaxy's gas supply. Coupled outflows, on the other hand, become \textit{more} effective because they are able to push more of a halo's baryons out to larger radii. Paper II builds upon those findings to explain how different cosmological simulations can successfully produce similar galaxy populations while making starkly different predictions for CGM properties.

\section{Summary}

This paper has presented a new regulator model for galaxy evolution that prioritizes simplicity and generality over complexity and detail. Its core is an accounting system---the minimalist regulator model---that tracks \textit{all} of the mass and energy associated with a halo's baryons, regardless of where they are located (\S \ref{sec:Dynamics}). That approach enables the model to represent expansion and contraction of a galaxy's circumgalactic atmosphere driven by imbalances between feedback heating and radiative cooling.

Here are the most important points that emerge from our analysis of that simple model:
\begin{enumerate}
    
    \item Supernova feedback's effects on galaxy evolution depend on whether the outflows that supernovae generate are \textit{coupled} to the CGM (\S \ref{sec:Foundation}). The minimalist regulator model applies to coupled outflows that share their energy with the rest of a galaxy's atmosphere.
    
    \item Increasing the mass loading parameter $\eta_M$ in the minimalist regulator model \textit{reduces} the net atmospheric energy change coming from the supernova feedback loop (\S \ref{sec:FlipSide}). That happens because the specific energy of a galactic wind ($\varepsilon_{\rm fb} = \eta_E \varepsilon_{\rm SN} / \eta_M$) is smaller if a given amount of feedback energy is spread over a greater gas mass.

    \item Whether or not a galaxy's atmosphere expands or contracts in response to supernova feedback depends on how the atmosphere's radiative cooling rate compares with the rate at which supernova feedback adds energy to the atmosphere (\S \ref{sec:Asymptotic}). The outcome hinges on whether the specific energy of gas returning to the atmosphere ($\varepsilon_{\rm fb}$) exceeds the specific energy it had when it first accreted onto the halo ($\varepsilon_{\rm acc}$). Large values of the mass loading parameter that reduce $\varepsilon_{\rm fb}$ below $\varepsilon_{\rm acc}$ result in atmospheric contraction. Smaller proportions of mass loading result in atmospheric expansion.

    \item Expanding galactic atmospheres asymptotically approach an equilibrium state in which supernova energy input keeps the atmosphere's specific energy constant as cosmological accretion adds atmospheric mass (\S \ref{sec:Expansion}). A galaxy's star formation rate in that asymptotic equilibrium state is
    \begin{equation}
        \dot{M_*} 
            \approx \left( 
                    \frac {\xi v_{\rm c}^2}
                    {\eta_E \varepsilon_{\rm SN}
                        + \xi v_{\rm c}^2}
                        \right)
                    \dot{M}_{\rm acc}
    \end{equation}
    in which $\xi$ is a model dependent dimensionless factor of order unity.
    The stellar baryon fraction of the galaxy's halo is \textit{insensitive} to $\eta_M$ because the atmospheric \textit{energy} input coming from supernova feedback is more consequential than ejection of galactic gas.

    \item Contracting galactic atmospheres can come into equilibrium, at least in principle, if supernova heating somehow manages to balance radiative cooling of the atmosphere (\S \ref{sec:Contraction}). The halo's equilibrium star formation rate then depends on the atmosphere's radiative cooling rate. However, such a state of balance may be precarious because of the complex dependence of radiative cooling on atmospheric structure and enrichment.


    \item Galactic outflows that do not share their energy with the rest of the CGM, which we have called \textit{uncoupled outflows}, cannot suppress the gas supply flowing from the CGM into a halo's central galaxy. They must regulate star formation in that galaxy by ejecting gas from it. However, uncoupled outflows with large proportions of mass loading cannot escape the halo and fall back toward the central galaxy if $\varepsilon_{\rm fb} < \varepsilon_{\rm acc}$ (\S \ref{sec:Ballistic}). Recycling of ejected gas that returns to the central galaxy boosts its star formation rate. The long-term star formation rate of a galaxy with uncoupled outflows then increases as $\eta_M$ rises.


    
\end{enumerate}
Some of these findings may seem obvious in retrospect, but they were not all obvious to the authors of this paper when we began to write it and are not universally appreciated within the community studying supernova feedback. Therein lie the benefits of a simple model.

Paper II discusses how the results of this paper help to explain differences among the CGM predictions made by current cosmological simulations. Future papers in this series will add more astrophysical content to the framework.\footnote{The next paper in the series (Pandya et al., in preparation) will call it the \textsc{ExpCGM} framework, short for \textit{Expandable CircumGalactic Medium}.} Our objective will be to learn how various assumptions about the structure and composition of galactic atmospheres and their halos affect model predictions for both galaxy evolution and observable features of the CGM. The framework's most distinctive feature is its focus on expansion/contraction rather than heating/cooling. This paper has tried to explain why expansion and contraction of a galaxy's atmosphere are more fundamental to the feedback loop than balance between heating and cooling.

Meanwhile, we invite the community to use the minimalist regulator model to interpret both observations and simulations of relationships between galaxy evolution and the CGM. In particular, we are hoping that phase-plane analyses like the idealized one shown in Figure \ref{fig:phaseplane_plots_10} will help to reveal how the star-formation histories of galaxies respond to the impact of feedback on their atmospheres.\footnote{Future versions of these phase-plane plots may feature a logarithmic axis showing $1 - f_{\rm CGM} = (M_* + M_{\rm ISM}) / f_{\rm b} M_{\rm halo}$ instead of a linear axis showing $f_{\rm CGM}$, so that results for low-mass halos with $f_* \ll 1$ are easier to visualize.}

\begin{acknowledgements}

GMV acknowledges support from the NSF through grant AAG-2106575 and also benefited from the \textit{Turbulence in Astrophysical Environments} program, supported in part by grant NSF PHY-2309135 to the Kavli Institute for Theoretical Physics (KITP).
NASA provides support for VP through the NASA Hubble Fellowship grant HST-HF2-51489 awarded by the Space Telescope Science Institute, which is operated by the Association of Universities for Research in Astronomy, Inc., for NASA, under contract NAS5-26555.  GLB acknowledges support from the NSF (AST-2108470, ACCESS PHY2400043, AST-2307419), NASA TCAN award 80NSSC21K1053, the Simons Foundation (grant 822237) and the Simons Collaboration on Learning the Universe. The Flatiron Institute is a division of the Simons Foundation. MD is grateful for partial support of this work from NASA award NASA-80NSSC22K0476. BDO's contribution was supported by Chandra Grant TM2-23004X 

\end{acknowledgements}

\appendix

\section{Descriptions of Symbols}

To help readers keep track of the many symbols used in the paper, Table \ref{tab:Glossary} provides a glossary, including the place where each symbol first appears. Also, the following list groups some subsets of those symbols into categories.

\begin{itemize}

    \item \textbf{Baryonic Energy Accounting:} $E_{\rm CGM}$, $E_\varphi$, $E_{\rm th}$, $E_{\rm nt}$ 

    \item \textbf{Baryonic Mass Accounting:} $M_{\rm acc}$, $M_{\rm CGM}$, $M_{\rm ISM}$, $M_*$ 

    \item \textbf{Baryonic Mass Exchange Rates:} $\dot{M}_{\rm acc}$, $\dot{M}_{\rm CGM}$, $\dot{M}_{\rm in}$, $\dot{M}_{\rm ISM}$, $\dot{M}_*$ 

    \item \textbf{Cosmological Halo Description:} $M_{\rm halo}$, $r_0$, $v_{\rm c}$, $\varphi$, $\dot{\varphi}$

    \item \textbf{Equilibrium State Analysis:} $f_{*,{\rm asy}}$, $\delta f_*$, $\varepsilon_{\rm eq}$, $\xi$

    \item \textbf{Galactic Atmosphere Model Description:} $\dot{E}_{\rm in}$, $\dot{E}_{\rm rad}$, $f_{\rm CGM}$, $f_{\rm in}$, $\dot{M}_{\rm in}$, $\varepsilon_{\rm CGM}$, $\varepsilon_{\rm in}$, $\varepsilon_{\rm loss}$, $\varepsilon_{\rm rad}$

    \item \textbf{Minimalist Regulator Model Parameters:} $t_{\rm SF}$, $\eta_E$, $\varepsilon_{\rm SN}$, $\eta_M$

    \item \textbf{Recycling Model Description:} $\dot{M}_{\rm in,0}$, $N_{\rm cyc}$, $t_{\rm cyc}$ 
    
\end{itemize}

\begin{table*}
    \centering
    \caption{Glossary of Symbols}
    \begin{tabular}{ccc}
       \hline \hline
       \textbf{Symbol} & \textbf{Description} & \textbf{First Used} \\
       \hline
         $E_{\rm CGM}$ 
         & Total CGM energy: $E_\varphi + E_{\rm th} + E_{\rm nt}$
         & Eq. \ref{eq:EnergyBudget}
         \\
         $\dot{E}_{\rm fb}$ 
         & Feedback energy output rate ($\eta_E \varepsilon_{\rm SN} \dot{M}_*$ for SN feedback)
         & Eq. \ref{eq:EnergySourcesSinks}
         \\
         $\dot{E}_{\rm in}$ 
         & Rate at which CGM loses energy as baryons move from CGM into ISM
         & Eq. \ref{eq:EnergySourcesSinks}
         \\
         $E_{\rm nt}$ 
         & Non-thermal CGM energy (including kinetic energy)
         & Eq. \ref{eq:EnergyBudget}
         \\
         $\dot{E}_{\rm rad}$ 
         & Radiative energy loss rate from the CGM 
         & Eq. \ref{eq:EnergySourcesSinks}
         \\
         $E_{\rm th}$ 
         & Thermal CGM energy
         & Eq. \ref{eq:EnergyBudget}
         \\
         $E_\varphi$ 
         & Gravitational potential energy of the CGM
         & Eq. \ref{eq:EnergyBudget}
         \\
         $f_{\rm b}$ 
         & Cosmic baryon mass fraction
         & \S \ref{sec:Foundation}
         \\
         $f_{\rm CGM}$ 
         & Fraction of accreted baryons in CGM: $M_{\rm CGM} / M_{\rm acc}$ 
         & Eq. \ref{eq:GasSupply}
         \\
         $f_{\rm in}$ 
         & Dimensionless function describing how $\dot{M}_{\rm in}$ depends on $\varepsilon_{\rm CGM}$
         & Eq. \ref{eq:GasSupply}
         \\
         $f_*$ 
         & Fraction of a halo's baryons in stars
         & \S \ref{sec:Introduction}
         \\
         $f_{*,{\rm asy}}$ 
         & Asymptotic stellar baryon fraction: $\dot{M}_{\rm in}/(1+\eta_M)\dot{M}_{\rm acc} $
         & Eq. \ref{eq:fstar_asy}
         \\
         $M_{\rm acc}$ 
         & Total mass of accreted baryons
         & Eq. \ref{eq:MassBudget}
         \\
         $M_{\rm CGM}$ 
         & Total CGM mass (including baryons pushed beyond $R_{\rm halo})$ 
         & Eq. \ref{eq:MassBudget}
         \\
         $M_{\rm halo}$ 
         & Mass of a galaxy's cosmological halo 
         & \S \ref{sec:Introduction}
         \\
         $M_{\rm ISM}$ 
         & Mass of a galaxy's interstellar medium  
         & Eq. \ref{eq:MassBudget}
         \\
         $\dot{M}_{\rm in}$ 
         & Gas supply rate at which baryons flow from CGM into ISM
         & Eq. \ref{eq:Mdot_ISM}
         \\
         $\dot{M}_{\rm in,0}$ 
         & Rate at which baryons flow from CGM into ISM for the first time
         & \S \ref{sec:Ballistic}
         \\
         $\dot{M}_{\rm wind}$ 
         & Mass outflow rate of galactic wind 
         & \S \ref{sec:Introduction}
         \\
         $M_*$ 
         & Central galaxy's stellar mass  
         & Eq. \ref{eq:MassBudget}
         \\
         $\dot{M}_*$ 
         & Central galaxy's star formation rate  
         & \S \ref{sec:Introduction}
         \\
         $N_{\rm cyc}$ 
         & Number of times a halo's baryons cycle through the central galaxy 
         & \S \ref{sec:Foundation}
         \\
         $r_0$ 
         & Inner radius of CGM, at which $\varphi = 0$ 
         & Eq.\ref{eq:SIS_Potential}
         \\
         $R_{\rm closure}$ 
         & Closure radius within which baryon mass fraction equals $f_{\rm b}$
         & \S \ref{sec:Foundation}
         \\
         $R_{\rm halo}$ 
         & Halo (virial) radius 
         & \S \ref{sec:Foundation}
         \\
         $t$ 
         & Cosmic time 
         & \S \ref{sec:Foundation}
         \\
         $t_{\rm cyc}$ 
         & Recycling timescale for halo baryons 
         & \S \ref{sec:Foundation}
         \\
         $t_{\rm SF}$ 
         & Star formation timescale for ISM: $M_{\rm ISM} / \dot{M}_*$ 
         & \S \ref{sec:Minimalist}
         \\
         $v_{\rm c}$ 
         & Circular velocity of cosmological halo
         & \S \ref{sec:Foundation}
         \\
         $\delta f_*$ 
         & Difference between $\dot{M}_*/\dot{M}_{\rm acc}$ and $f_{*,{\rm asy}}$
         & \S \ref{sec:EquilibriumAnalysis}
         \\
         $\varepsilon_{\rm acc}$ 
         & Specific energy of accreting baryons (including potential energy) 
         & \S \ref{sec:Foundation}
         \\
         $\varepsilon_{\rm CGM}$ 
         & Specific energy of CGM: $E_{\rm CGM} / M_{\rm CGM}$ 
         & \S \ref{sec:Expansion}
         \\
         $\varepsilon_{\rm eq}$ 
         & Specific energy of CGM in an equilibrium state 
         & \S \ref{sec:EquilibriumAnalysis}
         \\
         $\varepsilon_{\rm fb}$ 
         & Specific energy of galactic wind before coupling with CGM 
         & \S \ref{sec:Foundation}
         \\
         $\varepsilon_{\rm in}$ 
         & Specific energy of baryons entering the central galaxy: $\dot{E}_{\rm in} / \dot{M}_{\rm in}$ 
         & Eq. \ref{eq:eps_in}
         \\
         $\varepsilon_{\rm loss}$ 
         & Specific energy loss associated with the galaxy's gas supply: $(\dot{E}_{\rm rad} + \dot{E}_{\rm in}) / \dot{M}_{\rm in}$
         & Eq. \ref{eq:eps_loss}
         \\
         $\varepsilon_{\rm rad}$ 
         & Specific \textit{radiative} energy loss associated with the galaxy's gas supply: $\dot{E}_{\rm rad} / \dot{M}_{\rm in}$
         & Eq. \ref{eq:eps_rad}
         \\
         $\varepsilon_{\rm SN}$ 
         & Specific supernova energy output from a stellar population
         & \S \ref{sec:Foundation}
         \\
         $\eta_E$ 
         & Energy loading parameter: $\dot{E}_{\rm fb} / \dot{M}_* \varepsilon_{\rm SN}$ 
         & \S \ref{sec:Foundation}
         \\
         $\eta_M$ 
         & Mass loading parameter:  $\dot{M}_{\rm wind} / \dot{M}_*$
         & \S \ref{sec:Introduction}
         \\
         $\xi$ 
         & Model-dependent dimensionless energy difference:  $\xi = (\varepsilon_{\rm eq} - \varepsilon_{\rm acc})/v_{\rm c}^2$ 
         & \S \ref{sec:EquilibriumAnalysis}
         \\
         $\tau$ 
         & Dimensionless gas depletion parameter
         & \S \ref{sec:Minimalist}
         \\
         $\varphi$ 
         & Gravitational potential of cosmological halo (with zero point at $r_0$)
         & Eq. \ref{eq:SIS_Potential}
         \\ 
         $\dot{\varphi}$ 
         & Rate of change in the gravitational potential 
         & Eq. \ref{eq:Edot_phi}
         \\ 
         \hline
    \end{tabular}
    \label{tab:Glossary}
    \newpage
\end{table*}

\bibliography{CGM}{}
\bibliographystyle{aasjournal}

\end{document}